\documentclass[]{tJOT2e}

\usepackage{graphicx}
\usepackage{epstopdf}

\begin{document}
\doi{10.1080/14685240YYxxxxxxx}
 \issn{1468-5248}
 \jvol{12} \jnum{00} \jyear{2011}

\markboth{N. Yokoi}{Journal of Turbulence}

\articletype{RESEARCH ARTICLE}

\title{{\itshape Modeling the turbulent cross-helicity evolution: Production, dissipation, and transport rates}}

\author{N. Yokoi$^{\rm a,b}$$^{\ast}$$^{\dag}$\thanks{$^\ast$Corresponding author. Email: nobyokoi@iis.u-tokyo.ac.jp}
\thanks{$^\dag$Guest researcher at the National Astronomical Observatory of Japan (NAOJ)}
\vspace{6pt}\\
\vspace{6pt}  $^{\rm a}${\em{Institute of Industrial Science, University of Tokyo\\ 
4-6-1, Komaba, Meguro-ku, Tokyo 153-8505, Japan}}\\
\vspace{6pt} $^{\rm b}${\em{Nordic Institute for Theoretical Physics (NORDITA)\\
Roslagstullsbacken 23, 106 91 Stockholm, Sweden}}\\
\vspace{6pt}\received{23rd February, 2010, accepted 14th May, 2011} }

\makeatletter
\input subeqn.sty
\input subeqna.sty
\input fancyhdr.sty
\makeatother
%
%

\pagestyle{fancy}
\renewcommand{\headrulewidth}{0pt}
\lhead[\fancyplain{}{\rm\thepage}]{\fancyplain{Journal of Turbulence\\ TSFP-6
focus issue}{}}
\chead[\fancyplain{}{\it N. Yokoi}]{\fancyplain{}{Turbulent cross-helicity
evolution}}
\rhead[\fancyplain{}{}]{\fancyplain{}{\rm\thepage}}
\lfoot{}
\cfoot{}
\rfoot{}

\maketitle

\begin{abstract}\noindent
	It has been recognized that the turbulent cross helicity (correlation between
the velocity and magnetic-field fluctuations) can play an important role in
several magnetohydrodynamic (MHD) plasma phenomena such as the global
magnetic-field generation, turbulence suppression, etc. Despite its relevance
to the cross-helicity evolution, little attention has been paid to the
dissipation rate of the turbulent cross helicity, $\varepsilon_W$. In this paper,
we consider the model expression for the dissipation rate of the turbulent cross
helicity. In addition to the algebraic model, an evolution equation of $\varepsilon_W$ is proposed on the basis of the statistical analytical theory of inhomogeneous turbulence. A turbulence model with the modeling of $\varepsilon_W$ is applied to
the solar-wind turbulence. Numerical results on the large-scale
evolution of the cross helicity is compared with the satellite observations. It is
shown that, as far as the solar-wind application is concerned, the simplest
possible algebraic model for $\varepsilon_W$ is sufficient for elucidating the
large-scale spatial evolution of the solar-wind turbulence. Dependence of the
cross-helicity evolution on the large-scale velocity structures such as velocity
shear and flow expansion is also discussed.

\vspace{6pt}

\bigskip

\end{abstract}

\section{Introduction}\label{sec:level1}
	In the magnetohydrodynamic (MHD) turbulent flow at high magnetic Reynolds
number ($Rm \gg 1$), magnetic fields are considered to be frozen in plasmas,
and move with the flow.\cite{alf1950} In such a flow, the induced magnetic
field is often much larger than the originally imposed field. Besides, MHD waves
such as the Alfv\'{e}n wave are considered to exist ubiquitously. The cross
helicity, defined by the correlation between the velocity $\bf{u}$ and magnetic
field $\bf{b}$, is a possible describer of such MHD turbulence properties.
Actually, the magnetic-field generation due to the turbulent cross helicity has
been
investigated.\cite{yos1990,yos1993,yok1996,yos1998,yos1999,yok1999,yos2000,yos2004}

	As is well known, the total amount of cross helicity $\int_V {{\bf{u}} \cdot
{\bf{b}}} dV$, as well as that of the MHD energy $\int_V
{({\bf{u}}^2 + {\bf{b}}^2) / 2}\ dV$, is an inviscid invariant of the MHD
equations. Because of this conservative property, the turbulent densities of the
MHD energy and cross-helicity, $K \equiv \langle { {\bf{u}}' {}^2 +
{\bf{b}}' {}^2 } \rangle / 2$ and $W \equiv \langle { {\bf{u}}' \cdot {\bf{b}}' }
\rangle$, may serve themselves as a good measure for characterizing the
statistical properties of MHD turbulence (${\bf{u}}'$: velocity fluctuation,
${\bf{b}}'$: magnetic-field fluctuation, $\langle \cdots \rangle$: ensemble
average).

	The evolution equations of $K$ and $W$ are similar in form, and their
mathematical structures are quite simple. The evolution of $K$ and $W$ are
determined by three constitutes: the production, dissipation, and transport
rates. Firstly, the production rate is expressed by the correlation of turbulence
fields coupled with the mean-field inhomogeneity. We note that the production
rate of turbulence quantities such as $\langle {{\bf{u}}' \cdot {\bf{b}}'} \rangle$ can be expressed exactly in the same as the counterpart of mean-field
quantities such as ${\bf{U}} \cdot {\bf{B}}$, but with the opposite sign
($\bf{U}$: the mean velocity, $\bf{B}$: the mean magnetic field). This makes our
interpretation possible that the drain of mean-field quantity gives rise to the
generation of turbulence counterpart. So, the production rate represents how a
quantity is supplied to turbulence by way of its cascading process (See Appendix~\ref{sec:appendixA}). Secondly, the dissipation rates of the turbulent MHD energy and cross helicity, $\varepsilon$
and $\varepsilon_W$, whose definitions will be given shortly in
Section~\ref{sec:level2}, represent the effects of molecular viscosity and magnetic
diffusivity coupled with the small-scale fluctuations. However, we stress the following point. The dissipation rates of the turbulent MHD energy and cross helicity, $\varepsilon$ and $\varepsilon_W$, can be considered from another aspect. In the intermediate range of turbulence, called the inertial range, the energy and cross helicity supplied from the energy-containing range compensate the energy and cross-helicity lost in the dissipation range. For this cascade picture of turbulence, the energy and cross-helicity transfer from lower to higher wavenumber ranges are most important quantities. In an equilibrium turbulence, $\varepsilon$ and $\varepsilon_W$ represent these transfer rates of $K$ and $W$, respectively. This makes the construction of the $\varepsilon$ and $\varepsilon_W$ equation possible as we show later. Finally, the transport
rates express the flux of a quantity that enters the fluid volume through the
boundary. The expression for the transport rates suggests in what situation the
quantities considered can be supplied to turbulence.

	Thanks to these clear-cut pictures associated with the evolution equation, the cross helicity (density) $W$, as well as the turbulent MHD energy (density) $K$, may play an important role in the turbulence modeling of MHD fluids. However, as compared with $K$ and other pseudoscalar turbulence quantities such as the turbulence kinetic and magnetic helicities, only a limited attention so far has been paid to the cross helicity. 
	
	In the context of homogeneous isotropic MHD turbulence, some important investigations have been made on the decaying rate of the cross helicity or $\varepsilon_W$. It was shown that if there is a prevailed sign of the cross helicity in the initial state, the system goes towards a dynamically aligned state. The cross helicity scaled by the MHD energy grows towards +1 or -1 depending on the initially prevailed sign of the cross helicity.\cite{dob1980a,dob1980b,gra1982,gra1983}

	In the context of inhomogeneous MHD turbulence, the cross helicity has been investigated mostly in the solar-wind research. By using spacecraft observations, detailed spectra of cross helicity have been examined.\cite{bel1971,rob1987,tum1995} In order to explain the large-scale behavior of the solar-wind turbulence, several models have been proposed.\cite{zho1990,tum1995} However, investigations related to the cross helicity are mostly concentrated on arguments of its production rate, and effects of large-scale inhomogeneities such as the mean velocity shear have been discussed. Matthaeus and coworkers have employed a kind of algebraic model for the cross-helicity dissipation rate.\cite{zho1990,mat2004,bre2005,bre2008} Adopting this algebraic model of $\varepsilon_W$, Usmanov {\it et al}.\ have recently performed a series of elaborated numerical simulations on the large-scale evolution of solar-wind turbulence.\cite{usm2011} However, generally speaking, arguments concerning the dissipation rate of $W$ are still far from sufficient. 

	Also in the context of the turbulence dynamo, the transport equation for the turbulent cross helicity has been considered, where an algebraic model for the cross-helicity dissipation rate has been proposed.\cite{yos1990,yos1993} Sur and Brandenburg wrote down an evolution equation for the cross-helicity effect, and argued the cross-helicity destruction in the context of quenching mechanism.\cite{sur2009}

	In order to examine the evolution of the turbulent cross helicity $W$, it is indispensable to properly estimate the dissipation rate of $W$, $\varepsilon_W$, as well as the cross-helicity production rate $P_W$. We address this problem; modeling the cross-helicity dissipation rate on the basis of a statistical analytical theory.

	Spacecraft observations of solar-wind turbulence have revealed detailed information on the large-scale behavior of turbulent statistical quantities, which includes the radial evolution of the cross helicity both in the low- and high-speed wind regions. Comparison of the satellite observations with the numerical simulation with the aid of a turbulence model provides a good test for the cross-helicity dissipation models. In this work, we will delve into the problem of cross-helicity dissipation modeling by using such comparisons.

	The organization of this paper is as follows. After briefly showing the
evolution equation of the cross helicity in Section~\ref{sec:level2}, we present the
exact equation of the cross-helicity dissipation rate
$\varepsilon_W$ in Section~\ref{sec:level3}. Then, we consider two candidates for
the model of $\varepsilon_W$ in Section~\ref{sec:level4}; one is the algebraic model
and another is a transport-equation model. These expressions are systematically derived with the aid of a statistical analytical theory of inhomogeneous turbulence. Some features of the adopted turbulence model, which is an expansion of the hydrodynamic $k - \epsilon$-type one-point turbulence model in the engineering field, are noted in Section~\ref{sec:level5}. An application of the model to the solar wind is presented in Section~\ref{sec:level6}. A brief summary is given in Section~\ref{sec:level7}.

\section{Equation for the cross helicity}\label{sec:level2}
The fluctuation velocity and magnetic field, ${\bf{u}}'$ and ${\bf{b}}'$, in incompressible magnetohydrodynamic (MHD) flow are governed by
\begin{eqnarray}
	\left( {
	\frac{\partial}{\partial t}
	+ {\bf{U}} \cdot \nabla
	} \right) {\bf{u}}'
	&=& - \left( {{\bf{u}}' \cdot \nabla} \right) {\bf{U}}
	+ \left( {{\bf{B}} \cdot \nabla} \right) {\bf{b}}'
	+ \left( {{\bf{b}}' \cdot \nabla} \right) {\bf{B}}
	\nonumber\\
	& & - \left( {{\bf{u}}' \cdot \nabla} \right) {\bf{u}}'
	- \left( {{\bf{b}}' \cdot \nabla} \right) {\bf{b}}'
	+ \nabla \cdot \mbox{\boldmath${\cal{R}}$}
	- \nabla p'_{\rm{M}}
	+ \nu \nabla^2 {\bf{u}}'
	+ {\bf{f}}',
	\label{eq:fluct_u_eq}
\end{eqnarray}
\begin{eqnarray}
	\left( {
	\frac{\partial}{\partial t}
	+ {\bf{U}} \cdot \nabla
	} \right) {\bf{b}}'
	&=& - \left( {{\bf{u}}' \cdot \nabla} \right) {\bf{B}}
	+ \left( {{\bf{B}} \cdot \nabla} \right) {\bf{u}}'
	+ \left( {{\bf{b}}' \cdot \nabla} \right) {\bf{U}}
	\nonumber\\
	& & - \left( {{\bf{u}}' \cdot \nabla} \right) {\bf{b}}'
	- \left( {{\bf{b}}' \cdot \nabla} \right) {\bf{u}}'
	- \nabla \times {\bf{E}}_{\rm{M}}
	+ \lambda \nabla^2 {\bf{b}}'
	\label{eq:fluct_b_eq}
\end{eqnarray}
with the solenoidal conditions
\begin{equation}
		\nabla \cdot {\bf{u}}'
		= \nabla \cdot {\bf{b}}'
		= 0
	\label{eq:fluct_sol_cond}
\end{equation}
($\nu$: kinematic viscosity, $\lambda$: magnetic diffusivity). The primed quantities denote the deviations from the ensemble average $\langle \cdots \rangle$ as
\begin{equation}
		\varphi = \overline{\varphi} + \varphi',\;\; 
		\overline{\varphi} = \langle {\varphi} \rangle
\end{equation}
with
\begin{subequations}
\begin{equation}
		\varphi = \left( {
			{\rho, \bf{u}}, \mbox{\boldmath$\omega$}, {\bf{b}}, {\bf{j}}, p, 
			p_{\rm{M}}, {\bf{f}}
		} \right),
\end{equation}
\begin{equation}
		\overline{\varphi} = \left( {
			\overline{\rho}, {\bf{U}}, \mbox{\boldmath$\Omega$}, {\bf{B}}, {\bf{J}},
			P, P_{\rm{M}}, {\bf{F}}
		} \right),
\end{equation}
\begin{equation}
		\varphi' = \left( {
			\rho', {\bf{u}}', \mbox{\boldmath$\omega$}', {\bf{b}}', {\bf{j}}', 
			p', p'_{\rm{M}}, {\bf{f}}'
		} \right).
\end{equation}
\end{subequations}
Here, $\rho$ is the density, $\mbox{\boldmath$\omega$} (= \nabla \times {\bf{u}})$ the vorticity, ${\bf{j}} (= \nabla \times {\bf{b}})$ the electric-current density, $p$ the gas pressure, $p_{\rm{M}} (= p + {\bf{b}}^2 /2)$ the MHD pressure, and ${\bf{f}}$ the external force. Note that the magnetic field etc.\ are measured in the Alfv\'{e}n-speed unit. They are related to the ones measured in the original unit (asterisked) as
\begin{equation}
	{\bf{b}} = \frac{{\bf{b}}_\ast}{ (\mu_0 \rho)^{1/2}},\;\;\;
	{\bf{j}} = \frac{{\bf{j}}_\ast}{ (\rho/ \mu_0)^{1/2}},\;\;\;
	{\bf{e}} =  \frac{{\bf{e}}_\ast}{ (\mu_0 \rho)^{1/2}},\;\;\;
	p = \frac{p_\ast}{\rho}
	\label{eq:alfven_unit}
\end{equation}
($\mu_0$: magnetic permeability).

	The Reynolds stress $\mbox{\boldmath${\cal{R}}$}$ and the turbulent electromotive force ${\bf{E}}_{\rm{M}}$ represent the turbulence effects on the mean field. They are defined by
\begin{equation}
		{\cal{R}}^{\alpha\beta} 
		\equiv \left\langle {
		u'{}^\alpha u'{}^\beta - b'{}^\alpha b'{}^\beta 
		} \right\rangle,
	\label{eq:Reynolds_strss_def}
\end{equation}
\begin{equation}
		{\bf{E}}_{\rm{M}} 
		\equiv \left\langle{ {\bf{u}}' \times {\bf{b}}' }\right\rangle,
	\label{eq:emf_def}
\end{equation}
respectively.

	In order to describe the properties of turbulence, we consider several turbulent quadratic statistical quantities. Among them, the turbulent MHD energy $K$ and the turbulent cross helicity $W$ are defined by
\begin{equation} 
		K \equiv \frac{1}{2} \langle{{\bf{u}}'{}^2 + {\bf{b}}'{}^2}\rangle,
	\label{eq:K_def}
\end{equation}
\begin{equation}
		W \equiv \langle {{\bf{u}}' \cdot {\bf{b}}'} \rangle.
	\label{eq:W_def}
\end{equation}
From Equations~(\ref{eq:fluct_u_eq}) and (\ref{eq:fluct_b_eq}), the evolution equations of $K$ and $W $ are rightly obtained as
\begin{equation}
		\frac{DG}{Dt}
		\equiv \left( {
		\frac{\partial}{\partial t} + {\bf{U}} \cdot \nabla 
		}\right) G
		= P_G - \varepsilon_G + T_G
	\label{eq:transport_eq_G}
\end{equation}
with $G = (K, W)$. Here, $P_G$,
$\varepsilon_G$, and $T_G$ are the production, dissipation, and transport
rates of the turbulent statistical quantity $G$. They are defined by
\begin{subequations}\label{eq:K_eq_const}
\begin{equation}
		P_K = - {\cal{R}}^{ab} \frac{\partial U^a}{\partial x^b}
		- {\bf{E}}_{\rm{M}} \cdot {\bf{J}},
	\label{eq:P_K_def}
\end{equation}
\begin{equation}
		\varepsilon_K 
		= \nu \left\langle { \left( {
		\frac{\partial u'{}^a}{\partial x^b}
		} \right)^2 } \right\rangle
		+ \lambda \left\langle { \left( {
		\frac{\partial b'{}^a}{\partial x^b}
		} \right)^2 } \right\rangle
		\equiv \varepsilon,
	\label{eq:eps_K_def}
\end{equation}
\begin{equation}
		T_K
		 ={\bf{B}} \cdot  \nabla W 
		 - \nabla \cdot \left\langle {
		\left( { 
			\frac{{\bf{u}}'{}^2 + {\bf{b}}'{}^2}{2} + p'_{\rm{M}}
		} \right) {\bf{u}}'
		+ \left( {{\bf{u}}' \cdot {\bf{b}}'} \right) {\bf{b}}' 
		} \right\rangle
		+ \left\langle {{\bf{f}}' \cdot {\bf{u}}'} \right\rangle,
	\label{eq:T_K_def}
\end{equation}
\end{subequations}
\begin{subequations}\label{eq:W_eq_const}
\begin{equation}
		P_W
		= - {\cal{R}}^{ab} \frac{\partial B^a}{\partial x^b}
		- {\bf{E}}_{\rm{M}} \cdot {\mbox{\boldmath$\Omega$}},
	\label{eq:P_W_def}
\end{equation}
\begin{equation}
		\varepsilon_W
		= (\nu + \lambda) \left\langle {
		\frac{\partial u'{}^a}{\partial x^b} 
		\frac{\partial b'{}^a}{\partial x^b}
		} \right\rangle,
	\label{eq:eps_W_def}
\end{equation}
\begin{equation}
		T_W = {\bf{B}} \cdot \nabla K 
		- \nabla \cdot \left\langle {
		\left( {{\bf{u}}' \cdot {\bf{b}}'} \right) {\bf{u}}'
		- \left( { 
			\frac{{\bf{u}}'{}^2 + {\bf{b}}'{}^2}{2} - p'_{\rm{M}}
		} \right) {\bf{b}}' 
		} \right\rangle
		+ \left\langle {{\bf{f}}' \cdot {\bf{b}}'} \right\rangle.
	\label{eq:T_W_def}
\end{equation}
\end{subequations}

	As was already mentioned in Section~\ref{sec:level1} and Appendix~\ref{sec:appendixA}, the production rates $P_K$
and $P_W$ represent the supply of $K$ and $W$, respectively, due to turbulence
cascade. We see from the first term of Equation~(\ref{eq:P_K_def}) that the turbulent
energy is sustained by the mean velocity shear $\partial U^a / \partial x^b$. The
second or ${\bf{J}}$-related term of Equation~(\ref{eq:P_K_def}) is related to the Joule
dissipation of MHD turbulence. Equation~(\ref{eq:P_W_def}) shows that
inhomogeneities of the mean fields coupled with the fluctuation correlations are
essential for $P_W$. One is the shear of the large-scale magnetic field,
$\partial B^a / \partial x^b$, coupled with the Reynolds stress
$\mbox{\boldmath$\cal R$}$:
\begin{equation}
		P_{W\rm{R}} = - {\cal{R}}^{ab} \frac{\partial B^a}{\partial x^b}.
	\label{eq:PwR_def}
\end{equation}
The other is the large-scale vortical motion $\mbox{\boldmath$\Omega$}$ coupled
with the turbulent electromotive force
${\bf{E}}_{\rm{M}}$:
\begin{equation}
		P_{W\rm{E}} = - {\bf{E}}_{\rm{M}} \cdot \mbox{\boldmath$\Omega$}.
	\label{eq:PwE_def}
\end{equation}
Finite positive (or negative) values of $P_{W\rm{R}}$ and $P_{W\rm{E}}$ infer the generation of positive (or negative) turbulence cross helicity $W$ by way of cross-helicity cascade. The role of the cross-helicity production have been argued mainly in the context of turbulent dynamo and turbulent transport suppression. 

	Other important mechanisms that possibly supply the cross helicity to
turbulence come from the first and last terms of Equation~(\ref{eq:T_W_def}): 
\begin{equation}
		T_{W\rm{K}} = {\bf{B}} \cdot \nabla K,
	\label{eq:TwK_def}
\end{equation}
\begin{equation}
		T_{W\rm{F}} = \left\langle {{\bf{f}}' \cdot {\bf{b}}'} \right\rangle.
	\label{eq:TwF_def}
\end{equation}
Equation~(\ref{eq:TwK_def}) shows that the inhomogeneity of $K$ along the large-scale magnetic field $\bf{B}$ may contribute to the supply of the cross helicity.
We often meet such situations in astrophysical phenomena where inhomogeneous
turbulent plasma is threaded through by the ambient magnetic fields.

	On the other hand, Equation~(\ref{eq:TwF_def}) represents the turbulent cross-helicity generation due to the coupling of the external forcing and magnetic fluctuation. As we see in the evolution equation of the mean cross helicity ${\bf{U}} \cdot {\bf{B}}$ [Equation~(\ref{eq:T_WMB_def}) in Appendix~\ref{sec:appendixA}], the coupling of the (mean) external forcing with the large-scale magnetic field provides the mean cross helicity. Actually, such cross-helicity generation due to an external forcing plays a crucial role in the dynamo action or magnetic-field generation mechanism in a generalized Arnold--Beltrami--Childress flow called the Archonitis flow.\cite{sur2009} In this sense, analysis of the forcing effect on the evolution of the cross helicity is very important. It may affect the expression of the cross-helicity flux, just as the body-force effects such as the buoyancy, frame rotation, etc.\ change the scalar transport. However, this topic is beyond the scope of this paper.

	Unlike the production rates $P_K$ [Equation~(\ref{eq:P_K_def})] and $P_W$
[Equation~(\ref{eq:P_W_def})] and the first or ${\bf{B}}$-related terms of the
transport rates [Equations~(\ref{eq:T_K_def}) and (\ref{eq:T_W_def})], the definitions
of the dissipation rates, $\varepsilon$ [Equation~(\ref{eq:eps_K_def})] and
$\varepsilon_W$ [Equation~(\ref{eq:eps_W_def})], do not directly contain any mean
fields, but are expressed by the combination of the molecular viscosity $\nu$ and
magnetic diffusivity $\lambda$ with the inhomogeneity of the fluctuation fields.
So, we need some specific treatments on them. We consider the cross-helicity
dissipation rate in the following sections.

\section{Equation for the cross-helicity dissipation rate}\label{sec:level3}
			From the equations of the fluctuation velocity and magnetic field, we
construct the exact equation for the dissipation rate of the cross helicity,
$\varepsilon_W$ [Equation~(\ref{eq:eps_W_def})], as
\begin{eqnarray}
	\lefteqn{
	\frac{D \varepsilon_W}{Dt}
	\equiv \left( {
	\frac{\partial}{\partial t} + {\bf{U}} \cdot \nabla
	} \right) \varepsilon_W 
}\nonumber\\
& & \hspace{-15pt}= (\nu + \lambda) \left\langle {
	\frac{\partial u'{}^a}{\partial x^c}
	\frac{\partial b'{}^b}{\partial x^c}
	- \frac{\partial b'{}^a}{\partial x^c}
	\frac{\partial u'{}^b}{\partial x^c}
	} \right\rangle 
	\frac{\partial U^a}{\partial x^b}
\nonumber\\
& & \hspace{-15pt} + (\nu + \lambda) \left\langle {
	\frac{\partial u'{}^a}{\partial x^c} b'{}^b
	- u'{}^b \frac{\partial b'{}^a}{\partial x^c}
	} \right\rangle 
	\frac{\partial^2 U^a}{\partial x^b \partial x^c}
\nonumber\\
& & \hspace{-15pt} + (\nu + \lambda) \left\langle {
	\frac{\partial u'{}^b}{\partial x^a}
	\frac{\partial^2 u'{}^b}{\partial x^a \partial x^c}
	+ \frac{\partial b'{}^b}{\partial x^a}
	\frac{\partial^2 b'{}^b}{\partial x^a \partial x^c}
	} \right\rangle B^c
\nonumber\\
& & \hspace{-15pt} + (\nu + \lambda) \left\langle {
	\frac{\partial u'{}^c}{\partial x^a}
	\frac{\partial u'{}^c}{\partial x^b}
	- \frac{\partial b'{}^c}{\partial x^a}
	  \frac{\partial b'{}^c}{\partial x^b}
	} \right\rangle \frac{\partial B^a}{\partial x^b}
\nonumber\\
& & \hspace{-15pt} - (\nu + \lambda) \left\langle {
	\frac{\partial u'{}^a}{\partial x^c}
	\frac{\partial u'{}^b}{\partial x^c}
	- \frac{\partial b'{}^a}{\partial x^c}
	  \frac{\partial b'{}^b}{\partial x^c}
} \right\rangle \frac{\partial B^a}{\partial x^b}
\nonumber\\
& & \hspace{-15pt} - (\nu + \lambda) \left\langle {
	\frac{\partial u'{}^a}{\partial x^c} u'{}^b
	- \frac{\partial b'{}^a}{\partial x^c} b'{}^b
	} \right\rangle 
	\frac{\partial^2 B^a}{\partial x^b \partial x^c}
\nonumber\allowdisplaybreaks\\
& & \hspace{-15pt} - (\nu + \lambda) \left\langle {
	\frac{\partial u'{}^b}{\partial x^a}
	\frac{\partial u'{}^c}{\partial x^a}
	\frac{\partial b'{}^b}{\partial x^c}
	} \right\rangle
- (\nu + \lambda) \left\langle {
	\frac{\partial b'{}^b}{\partial x^a}
	\frac{\partial u'{}^c}{\partial x^a}
	\frac{\partial u'{}^b}{\partial x^c}
	} \right\rangle
\nonumber\\
& & \hspace{-15pt} + (\nu + \lambda) \left\langle {
	\frac{\partial u'{}^b}{\partial x^a}
	\frac{\partial b'{}^c}{\partial x^a}
	\frac{\partial b'{}^b}{\partial x^c}
	} \right\rangle
+ (\nu + \lambda) \left\langle {
	\frac{\partial b'{}^b}{\partial x^a}
	\frac{\partial b'{}^c}{\partial x^a}
	\frac{\partial b'{}^b}{\partial x^c}
	} \right\rangle
\nonumber\allowdisplaybreaks\\
& & \hspace{-15pt} - (\nu + \lambda) \left\langle {
	(u'{}^c \pm b'{}^c)
		\frac{\partial}{\partial x^c} 
		\left( {\frac{\partial u'{}^b}{\partial x^a}
				\frac{\partial b'{}^b}{\partial x^a}
		} \right)
	} \right\rangle
\nonumber\\
& & \hspace{-15pt} + (\nu + \lambda) \frac{\partial}{\partial x^c} 
	\left\langle {
		\frac{1}{2} b'{}^c \left[ {
			\frac{\partial}{\partial x^a}
			\left( {u'{}^b \pm b'{}^b} \right)
		} \right]^2
	} \right\rangle
\nonumber\allowdisplaybreaks\\
& & \hspace{-15pt} - (\nu + \lambda) \left\langle {
		\frac{\partial b'{}^b}{\partial x^a}
		\frac{\partial^2 p_{\rm{M}}'}{\partial x^a \partial x^b}
	}\right\rangle
+ (\nu + \lambda) 
	\frac{\partial^2}{\partial x^c \partial x^c}
	\varepsilon_W
\nonumber\\
& & \hspace{-15pt} - (\nu + \lambda) \frac{\partial}{\partial x^c} \left[{
	\nu \left\langle {
		\frac{\partial u'{}^b}{\partial x^a}
		\frac{\partial^2 b'{}^b}{\partial x^a \partial x^c}
	}\right\rangle
	+ \lambda \left\langle {
		\frac{\partial b'{}^b}{\partial x^a}
		\frac{\partial^2 u'{}^b}{\partial x^a \partial x^c}
	}\right\rangle
	}\right]
\nonumber\\
& & \hspace{-15pt} - (\nu + \lambda)^2 \left\langle{
	\frac{\partial^2 u'{}^b}{\partial x^a \partial x^c}
	\frac{\partial^2 b'{}^b}{\partial x^a \partial x^c}
	}\right\rangle
	\label{eq:eps_W_eq_exact}
\end{eqnarray}
(both upper or both lower signs should be chosen in the double signs). This equation, lacking the connection with a conservative law, has a
considerably complicated structure. This is in sharp contrast to the $K$ and $W$
equations [Equation~(\ref{eq:transport_eq_G})].

	In the hydrodynamic case with an electrically non-conducting
fluid, the energy dissipation rate
\begin{equation}
		\epsilon
		\equiv \nu \left\langle {
		\frac{\partial u'{}^a}{\partial x^b}
		\frac{\partial u'{}^a}{\partial x^b}
		} \right\rangle,
	\label{eq:eps_def_hd}
\end{equation}
as well as the turbulent energy $K_u \equiv \left\langle { {\bf{u}}'{}^2 }
\right\rangle / 2$, plays a central role in turbulence modeling.\cite{lau1972}
From the equation of the velocity fluctuation, we write the
$\epsilon$ equation exactly as
\begin{eqnarray}
	\lefteqn{
	\frac{D \epsilon}{Dt}
	\equiv \left( {
	\frac{\partial}{\partial t} + {\bf{U}} \cdot \nabla
	} \right) \epsilon
}\nonumber\\
& & = - 2 \nu \left\langle {
	\frac{\partial u'{}^b}{\partial x^a}
	\frac{\partial u'{}^c}{\partial x^a}
	\frac{\partial u'{}^b}{\partial x^c}
	} \right\rangle
- 2 \left\langle{ \left( {
	\nu \frac{\partial^2 u'{}^b}{\partial x^a \partial x^c}
	} \right)^2 }\right\rangle
\nonumber\\
& & - 2 \nu \left\langle {
	\frac{\partial u'{}^a}{\partial x^c}
	\frac{\partial u'{}^b}{\partial x^c}
	+ \frac{\partial u'{}^a}{\partial x^c}
	\frac{\partial u'{}^b}{\partial x^c}
	} \right\rangle 
	\frac{\partial U^a}{\partial x^b}
\nonumber\\
& &  - 2 \nu \left\langle {
	u'{}^b \frac{\partial u'{}^a}{\partial x^c}
	} \right\rangle 
	\frac{\partial^2 U^a}{\partial x^b \partial x^c}
\nonumber\\
& & + \frac{\partial}{\partial x^c} \left[{
	- \nu \left\langle { \left( {
		\frac{\partial u'{}^b}{\partial x^a}
		} \right)^2
	}\right\rangle
- 2 \nu \left\langle {
		\frac{\partial p'}{\partial x^b}
		\frac{\partial u'{}^a}{\partial x^b}
	}\right\rangle
	}\right]
\nonumber\\
& & + \nu 
	\frac{\partial^2 \epsilon}{\partial x^c \partial x^c}.
	\label{eq:eps_eq_exact}
\end{eqnarray}
The mathematical structure of this equation is also complicated because of the
lack of the connection with a conservative law.

	The energy dissipation $\epsilon$ itself is dominant at small scales. Using
this, the length and velocity scales are estimated as
\begin{equation}
		|{\bf{x}}| \sim \nu^{3/4} \epsilon^{-1/4},\;\;
		|{\bf{u}}'| \sim \nu^{1/4} \epsilon^{1/4},
	\label{eq:eps_scaling}
\end{equation}
respectively. Using Equation~(\ref{eq:eps_scaling}), we can
estimate each term in Equation~(\ref{eq:eps_eq_exact}). In the flow at high Reynolds number ($\nu \to 0$), two terms behaving as $O(\nu^{-1/2} \epsilon^{3/2})$ are dominant,
and these two terms should balance each other\cite{ten1972,yos1998}
\begin{equation}
	- 2 \nu \left\langle {
	\frac{\partial u'{}^a}{\partial x^b}
	\frac{\partial u'{}^a}{\partial x^c}
	\frac{\partial u'{}^b}{\partial x^c}
	} \right\rangle
\sim 2 \left\langle{ \left( { \nu
	\frac{\partial^2 u'{}^a}{\partial x^b \partial x^c}
	} \right)^2 }\right\rangle.
	\label{eq:dominant_eps_terms}
\end{equation}
In other word, in the modeling of the $\epsilon$ equation, it is of crucial
importance to properly estimate the two terms in
Equation~(\ref{eq:dominant_eps_terms}).

	In the hydrodynamic turbulence modeling, an empirical model equation for
$\epsilon$:
\begin{equation}
	\frac{D \epsilon}{Dt}
	\equiv \left( {
		\frac{\partial}{\partial t}
		+ {\bf{U}} \cdot \nabla
	} \right) \epsilon
	= C_{\epsilon 1} {} \frac{\epsilon}{k} P_{k}
	- C_{\epsilon 2} {} \frac{\epsilon}{k} \epsilon
	+ \nabla \left( {
		\frac{\nu_{\rm{T}}}{\sigma_\epsilon} \nabla \epsilon
	} \right)
	\label{eq:eps_eq_hd}
\end{equation}
was proposed and has been widely accepted as useful.

	Here, $C_{\epsilon 1}$, $C_{\epsilon 2}$, and $\sigma_{\epsilon}$ are model
constants. The values of these constants have been optimized through various
applications of the $k-\epsilon$ model. Usually, the values of
\begin{equation}
	C_{\epsilon 1} = 1.4,\;\;
	C_{\epsilon 2} = 1.9,\;\;
	\sigma_{\epsilon} = 1.0
	\label{eq:eps_model_const}
\end{equation}
are adopted.\cite{lau1972}

	Equation~(\ref{eq:eps_eq_hd}) is based on an empirical inference that the dissipation should be larger where the turbulence is larger, and is derived by
dimensional analysis. However, it is also pointed out that the system of
constants given by Equation~(\ref{eq:eps_model_const}) is not a unique combination for
the model constants because $\epsilon$ equation can take an infinite
number of self-similar states.\cite{rub2005}

	A similar argument using Equation~(\ref{eq:eps_scaling}) can be applied to the exact equation for the dissipation rate of the turbulent cross helicity
[Equation~(\ref{eq:eps_W_eq_exact})]. The difference between the molecular viscosity
$\nu$ and the magnetic diffusivity $\lambda$ is expressed by the magnetic Prandtl
number:
\begin{equation}
	Pm \equiv {\nu}/{\lambda}.
	\label{eq:mag_Pr_def}
\end{equation}
If we estimate each term in Equation~(\ref{eq:eps_W_eq_exact}) under the simplest
possible condition of $Pm = 1$, we have a dominant balance between the terms
expressed by
\begin{subequations}
\begin{equation}
	(\nu + \lambda) \left\langle {
	\frac{\partial u'{}^b}{\partial x^a}
	\frac{\partial u'{}^c}{\partial x^a}
	\frac{\partial b'{}^b}{\partial x^c}
	} \right\rangle
	\sim \nu^{-1/2} \varepsilon^{3/2},
\end{equation}
\begin{equation}
	(\nu + \lambda)^2 \left\langle{
	\frac{\partial^2 u'{}^b}{\partial x^a \partial x^c}
	\frac{\partial^2 b'{}^b}{\partial x^a \partial x^c}
	}\right\rangle
	\sim \nu^{-1/2} \varepsilon^{3/2}
\end{equation}
\end{subequations}
in the $\varepsilon_W$ equation at the high Reynolds number flow.

\section{Models for the dissipation of the turbulent cross
helicity}\label{sec:level4}

\subsection{\it\textbf{Algebraic model}}\label{sec:level4-1}
	As was mentioned in the previous section, the equation governing the
dissipation rate of $W$, $\varepsilon_W$, is very complicated. In such a
situation, the simplest possible model for $\varepsilon_W$ is the algebraic
approximation as follows.

	Using the turbulent MHD energy $K$ and its dissipation rate $\varepsilon$, we
construct a characteristic time scale of turbulence as
\begin{equation}
	\tau = {K}/{\varepsilon}.
	\label{eq:eddy_turnover_time}
\end{equation}
Other choices of time scale are possible. A large-scale magnetic field may alter
the characteristics of turbulence. In the case of MHD turbulence,
the Alfv\'{e}n time associated with the magnetic field may modulate time scale
of turbulence. This point will be referred to later at the end of
Section~\ref{sec:level5}.

	With the aid of the time scale of Equation~(\ref{eq:eddy_turnover_time}), the
dissipation rate of $W$ can be modeled as
\begin{equation}
		\varepsilon_W
		= C_W \frac{W}{\tau}
		= C_W \frac{\varepsilon}{K} W,
	\label{eq:algebraic_eps_w}
\end{equation}
where $C_W$ is the model constant. Namely, we consider the dissipation of the
turbulent cross helicity is proportional to the turbulent cross helicity divided
by the time scale.

	It is worth noting a mathematical constraint on the cross helicity. Namely, the
magnitude of the turbulent cross helicity $W$ is bounded by the magnitude of the
turbulent MHD energy $K$ as
\begin{equation}
		\frac{|W|}{K}
		= \frac{| \langle {\bf{u}}' \cdot {\bf{b}}' \rangle|}
		{\langle{\bf{u}}'{}^2 + {\bf{b}}'{}^2 \rangle / 2}
		\le 1.
	\label{eq:W_K_constraint}
\end{equation}
This relation constrains the value of $\varepsilon_W$. Actually, from
Equation~(\ref{eq:transport_eq_G}), the turbulent cross helicity scaled by the
turbulent MHD energy, $W/K$, is subject to
\begin{eqnarray}
		\lefteqn{
		\frac{D}{Dt} \frac{W}{K}
		= \frac{W}{K} \left( {
			\frac{1}{W} \frac{DW}{Dt} - \frac{1}{K} \frac{DK}{Dt}
		} \right)
		}\nonumber\\
		&=& \frac{W}{K} \left( {\frac{1}{W} P_W - \frac{1}{K} P_K} \right)
		- \frac{W}{K} \left( {
			\frac{1}{W} \varepsilon_W - \frac{1}{K} 	\varepsilon
		} \right)
		\nonumber\\
		&+& \frac{W}{K} \left( {\frac{1}{W} T_W - \frac{1}{K} T_K} \right).
	\label{eq:scaled_W_eq}
\end{eqnarray}

	Equation (\ref{eq:scaled_W_eq}) is a very general expression for the evolution equation of the scaled cross helicity, $W/K$, which should be satisfied with in any situations of turbulence flow. If we consider homogeneous turbulence, where the spatial variations of mean quantities vanish, we have neither production nor transport rate ($P_G = T_G = 0$). In such a case, Equation~(\ref{eq:scaled_W_eq}) is reduced to
\begin{equation}
	\frac{\partial}{\partial t} \frac{W}{K}
	= - \left( {
		\frac{1}{W} \varepsilon_W - \frac{1}{K} \varepsilon
	} \right) \frac{W}{K}.
	\label{eq:homo_WoverK_eq}
\end{equation}
This may give a constraint on the values of the dissipation rates of the turbulent MHD energy and cross helicity, $\varepsilon$ and $\varepsilon_W$. At least, Equation~(\ref{eq:homo_WoverK_eq}) provides us with a constraint for the turbulence modeling of $\varepsilon$ and $\varepsilon_W$ as we see in the following.

	If the coefficient of $W/K$ or parenthesized quantity in Equation~(\ref{eq:homo_WoverK_eq}) does not depend on $W/K$, which occurs naturally when we adopt simple algebraic models for $\varepsilon$ and $\varepsilon_W$ as below, in order for the magnitude of the scaled cross helicity to be bounded in time evolution, the parenthesized quantity should be positive. Then we have inequality
\begin{equation}
		\frac{|\varepsilon_W|}{\varepsilon} > \frac{|W|}{K}.
	\label{eq:W_K_constraint_basic}
\end{equation}
Provided that the difference between the time scales of $W$ and $K$ is independent of $W/K$, this is a necessary condition for $\varepsilon_W$ in relation to $\varepsilon$ arising from the upper-boundedness of $W$ [Equation~(\ref{eq:W_K_constraint})]. Actually, if we adopt simple algebraic models for $\varepsilon$ and $\varepsilon_W$ such as
\begin{equation}
		\varepsilon = \frac{K}{\tau},\;\;
		\varepsilon_W = C_W \frac{W}{\tau},
	\label{eq:simple_algeb_models}
\end{equation}
the parenthesized quantity in Equation~(\ref{eq:homo_WoverK_eq}) does not depend on $W/K$. Then we have inequality (\ref{eq:W_K_constraint_basic}), which gives a constraint on the model constant $C_W$ as will be seen later in Equation~(\ref{eq:C_w_condition_alg}) in Section~6.2.

	In the context of homogeneous MHD turbulence, since the pioneering work by Dobrowolny, Mageney \& Veltri\cite{dob1980a, dob1980b}, it had been considered that the initially prevailing cross helicity grows with time due to the nonlinear interaction towards the $+1$ (or $-1$) value of the scaled cross helicity (Also see \cite{gra1982,gra1983}). However, further investigation by Ting et al.\cite{tin1986} and Stribling \& Matthaeus\cite{str1991} showed that MHD turbulence has diverse possibilities for long time evolution, including either growth or reduction of $W/K$. In contrast to such homogeneous MHD turbulence, in inhomogeneous MHD turbulence with non-vanishing mean velocity shear, we have production, dissipation, and transport mechanisms of the turbulent cross helicity arising from or related to the inhomogeneity of mean fields. In the present work, we explore these mechanisms intrinsic to the inhomogeneities of MHD turbulence. There have been several works where the effects of mean velocity shear are incorporated into the evolution equations of turbulent quantities including the turbulent cross helicity. In this sense, this work is in the same line with \cite{mat2004,bre2005,bre2008}, and also \cite{yok2006,yok2007,yok2008}. However, in the present work, special emphasis is placed on the theoretical derivation of the turbulent cross-helicity dissipation-rate equation as will be shown in the following sections.

	Here, we had better remark on the notation of the scaled cross helicities. The turbulent cross helicity scaled by the turbulent MHD energy, $W/K$ [Equation~(\ref{eq:W_K_constraint})], is ``dynamically'' important in the context of turbulent dynamo etc. There is another scaled cross helicity
\begin{equation}
		\Gamma 
		= \frac{\langle {{\bf{u}}' \cdot {\bf{b}}'} \rangle}
			{ \sqrt{ \langle {{\bf{u}}'{}^2} \rangle \langle {{\bf{b}}'{}^2} \rangle}}
	\label{eq:align_meas}
\end{equation}	
that is ``kinematically'' or geometrically important, since it represents the degree of alignment of the velocity and magnetic-field vectors. In the solar-wind turbulence community, the former, $W/K$, is often referred to as ``$\sigma_{\rm{C}}$''. However, in the ``general'' turbulence community, this is not the case; Some use ``$\rho_{\rm{C}}$'' for this quantity, others do ``$\rho$'', and so on. Hence, in this paper, we confine ourselves to just mentioning these notation conventions. 

Interestingly, the directional alignment expressed by $\Gamma$ has shown to be robust rather than the behaviour of $W/K$.\cite{rob1992} This tendency is related to the fact that, unlike $\Gamma$, $W/K$ may dynamically change its value depending on how much differently the turbulent MHD energy and cross helicity, $K$ and $W$, are influenced by the large-scale shears. This point will be referred to later in Sec.~\ref{sec:level6-4}.

\subsection{\it\textbf{Model equation for the cross-helicity dissipation rate
($\varepsilon_W$ equation)}}\label{sec:level4-2}
	The equation of the $K$ dissipation rate, $\varepsilon$ equation, can be written as
\begin{equation}
		\frac{\partial \varepsilon}{\partial t}
		+ \left( {\bf{U} \cdot \nabla} \right) \varepsilon
		= C_{\varepsilon 1} \frac{\varepsilon}{K} P_K
		- C_{\varepsilon 2} \frac{\varepsilon}{K} \varepsilon
		+ \nabla \cdot \left( {
			\frac{\nu_K}{\sigma_{\varepsilon}} \nabla \varepsilon
		} \right).
	\label{eq:eps_eq_mhd}
\end{equation}
This is a direct expansion of the hydrodynamic (HD) or non-MHD turbulence energy dissipation $\epsilon$ equation~(\ref{eq:eps_eq_hd}). The model constants in Equation~(\ref{eq:eps_eq_mhd}) are same as the ones appearing in Equation~(\ref{eq:eps_eq_hd}). This is a consequence of the requirement that the $\varepsilon$ equation for MHD turbulence should be reduced to the $\epsilon$ equation for the HD in the limit of the vanishing magnetic field (${\bf{B}} = 0, {\bf{b}}' = 0$).

	We see from Equation~(\ref{eq:transport_eq_G}) that the equations of $W$ and $K$ are written in a similar form. So, it is natural to consider that the equation of the dissipation rate of $W$, $\varepsilon_W$, can be expressed in a form similar to the equation of the $K$ dissipation rate $\varepsilon$. Then we may consider the equation for $\varepsilon_W$ as
\begin{equation}
	\frac{{D\varepsilon_W}}{{Dt}} 
	= C_{W 1} \frac{\varepsilon}{K} P_W  
	- C_{W 2} \frac{\varepsilon}{K} \varepsilon_W 
	+ \nabla \cdot \left( {
		\frac{\nu_K}{\sigma_{\varepsilon W}} \nabla \varepsilon_W
	} \right),
	\label{eq:eps_W_model_eq}
\end{equation}
where $C_{W1}$, $C_{W2}$, and $\sigma_{\varepsilon W}$ are model constants.

	In contrast to the MHD energy, the cross helicity, defined by the correlation between the velocity and magnetic field, vanishes in the limit of the vanishing magnetic field. As this result, as far as the $\varepsilon_W$ equation is concerned, we can not make use of the knowledge accumulated in the history of HD turbulence modeling. To say nothing of the model constants $C_{W1}$, $C_{W2}$, and $\sigma_{\varepsilon W}$ appearing in Equation~(\ref{eq:eps_W_model_eq}), we have to consider the structure of $\varepsilon_W$ equation itself from more fundamental basis. For this purpose, in the following we will construct the equation for the cross-helicity dissipation rate on the analytical theoretical basis.

	From the theoretical analysis of inhomogeneous MHD turbulence, the cross-helicity density defined by $W \equiv \langle {{\bf{u}}' \cdot {\bf{b}}'} \rangle$ is expressed as
\begin{equation}
		W  = 2 I_0 \{Q_{ub}\}
		- I_0 \left\{ {
			G_{\rm{S}}, \frac{D}{Dt} \left( {Q_{ub} + Q_{bu}} \right)
		} \right\},
	\label{eq:W_from_TSDIA}
\end{equation}
where we have used abbreviated forms of spectral and time integrals:
\begin{subequations}\label{eq:abbrevs}
\begin{equation}
		I_n \left\{ {A} \right\}
		= \int {A(k,{\bf{x}}; \tau,\tau,t)} k^{2n} d{\bf{k}},
	\label{eq:abbrev1}
\end{equation}
\begin{equation}
		I_n \left\{ {A, B} \right\}
		= \int d{\bf{k}}\ k^{2n} \int_{-\infty}^\tau \!\!\!\!d\tau_1
		A(k,{\bf{x}}; \tau,\tau_1,t)
		B(k,{\bf{x}}; \tau,\tau_1,t).
	\label{eq:abbrev2}
\end{equation}
\end{subequations}
In Equation~(\ref{eq:W_from_TSDIA}), $G_{\rm{S}}$ is the Green's function, and $Q_{ub}$ and $Q_{bu}$ are the correlation functions related to the basic or lowest-order fields ${\bf{u}}_{\rm{B}}'$ and ${\bf{b}}_{\rm{B}}'$:
\begin{equation}
	\left\langle {
		{\bf{u}}'_{\rm{B}}\cdot{\bf{b}}'_{\rm{B}}
	} \right\rangle
	= 2 \int {Q_{ub}(k;\tau,\tau)} d{\bf{k}}.
	\label{eq:Qub_def}
\end{equation}
For the derivation of Equation~(\ref{eq:W_from_TSDIA}) and higher-order expression for $W$, see Appendix~\ref{sec:appendixB}. Suggestions from the higher-order expressions for the cross-helicity dissipation model are also presented in Appendix~{\ref{sec:appendixC}}.

	We assume that the correlation function and the Green's function in the inertial range are expressed as
\begin{equation}
	Q_{ub}(k,{\bf{x}};\tau,\tau',t)
	= \sigma_W(k,{\bf{x}};t)
	\exp\left[ {-\omega_W(k,{\bf{x}};t)|\tau - \tau'|} \right],
	\label{eq:spect_Qub}
\end{equation}
\begin{equation}
	G_{\rm{S}}(k,{\bf{x}};\tau,\tau',t)
	= \theta(\tau - \tau')
	\exp\left[ {-\omega_{\rm{S}}(k,{\bf{x}};t)(\tau - \tau')} \right],
	\label{eq:spect_Gs}
\end{equation}
where $\theta(\tau)$ is the Heaviside's step function that is 1 and 0 for $\tau > 0$ and $\tau<0$, respectively. Here, $\sigma_W$ is the power spectra of the turbulent cross helicity, and $\omega_W$ and $\omega_{\rm{S}}$ represent the frequencies or time scales of fluctuations. As for the spectra in the inertial range, we assume

\begin{equation}
	\sigma_W(k,{\bf{x}};t) 
	= \sigma_{W0} \varepsilon^{-1/3} \varepsilon_W({\bf{x}};t) k^{-11/3},
	\label{eq:sigmaW}
\end{equation}
and for the time scales,
\begin{equation}
	\omega_{\rm{S}}(k,{\bf{x}};t)
	= \omega_{\rm{S}0} \varepsilon^{1/3} k^{2/3} = \tau_{\rm{S}}^{-1},
	\label{eq:spect_omega_s}
\end{equation}
\begin{equation}
	\omega_{W}(k,{\bf{x}};t)
	= \omega_{W0} \varepsilon_W^{1/3} k^{2/3} = \tau_W^{-1}
	\label{eq:spect_omega_W}
\end{equation}
($\sigma_{W0}$, $\omega_{\rm{S}0}$, and $\omega_{W0}$ are numerical factors). Equation (\ref{eq:sigmaW}) arises from the assumption that the spectrum of the cross helicity is determined by the scale, energy and cross-helicity transfer rates; $k$, $\varepsilon$ and $\varepsilon_W$.

	Using Equations~(\ref{eq:spect_Qub}) and (\ref{eq:spect_Gs}), $W$ is calculated as
\begin{eqnarray}
		\lefteqn{
		W 
		= 2 \int d{\bf{k}}\ \varepsilon^{-1/3}({\bf{x}};t) 
			\varepsilon_W({\bf{x}};t) k^{-11/3}
		}\nonumber\\
		& & -2 \int d{\bf{k}} \left[ {
		\frac{1}{\omega_{\rm{S}} + \omega_{W}}
		\frac{D\sigma_W}{Dt}
		- \frac{\sigma_W}{(\omega_{\rm{S}} + \omega_W)^2}
		\frac{D\omega_W}{Dt}
		} \right].
	\label{eq:W_lower}
\end{eqnarray}
From the inertial-range forms (\ref{eq:sigmaW})-(\ref{eq:spect_omega_W}), this can be rewritten as
\begin{eqnarray}
	\lefteqn{
	W 
	= 4 \cdot 2\pi \varepsilon^{-1/3} \varepsilon_W 
	\int_{|{\bf{k}}| \ge k_{\rm{C}}} \!\!\!\!\!\!dk\ 
	k^{-5/3}
	}\nonumber\\
	& & - 4 \cdot 2 \pi \frac{\sigma_{W0}}{\omega_{\rm{sw}}\varepsilon_{\rm{sw}}^{1/3}} 
	\int_{|{\bf{k}}|\ge k_{\rm{C}}}\!\!\!\!\!\! dk\ 
	k^{4/3} \frac{D}{Dt} \left[ {
		\varepsilon^{-1/3}({\bf{x}};t) 
		\varepsilon_W({\bf{x}};t) 
		k^{-11/3}
	} \right]
	\nonumber\\
	& & + 4 \cdot 2 \pi \frac{\sigma_{W0} \omega_{W0}}
	{(\omega_{\rm{sw}}\varepsilon_{\rm{sw}}^{1/3})^2}
	\varepsilon^{-1/3} \varepsilon_W
		\int_{|{\bf{k}}|\ge k_{\rm{C}}}\!\!\!\!\!\! dk\ 
		k^{-3} \frac{D}{Dt} \left[ {
		\varepsilon_W^{1/3}({\bf{x}};t) k^{2/3}
		} \right]
	\label{eq:W_kC}
\end{eqnarray}
($k_{\rm{C}}$: cut-off wave number). Here, we have introduced a synthesized time scale $\tau_{\rm{SW}}$ defined by
\begin{equation}
	\frac{1}{\tau_{\rm{SW}}}
	= \frac{1}{\tau_{\rm{S}}} + \frac{1}{\tau_{W}}
	=  \left( {\omega_{\rm{S}0} \varepsilon^{1/3}
	+ \omega_{W0} \varepsilon_{W}^{1/3}} \right) k^{2/3}
	\equiv \omega_{\rm{sw}} \varepsilon_{\rm{sw}}^{1/3} k^{2/3}.
	\label{eq:eps_eff}
\end{equation}
We should note that $\omega_{\rm{sw}}$ and $\varepsilon_{\rm{sw}}$ will appear only in the combination of $\omega_{\rm{sw}} \varepsilon_{\rm{sw}}^{1/3}$; each of $\omega_{\rm{sw}}$ and $\varepsilon_{\rm{sw}}$ has no definite meaning. For simplicity of notation, hereafter we denotes
\begin{equation}
	A_W(\omega_{\rm{S}0}, \omega_{W0})
	\equiv \frac{\omega_{W0} \varepsilon_W^{1/3}} 
	{\omega_{\rm{sw}} \varepsilon_{\rm{sw}}^{1/3}}
	= \frac{\omega_{W0} \varepsilon_W^{1/3}} 
	{\omega_{\rm{S}0} \varepsilon^{1/3}
	+ \omega_{W0} \varepsilon_W^{1/3}}
	= \frac{\tau_{\rm{SW}}}{\tau_{W}}.
	\label{eq:A_epsepsW_def}
\end{equation}
In Equation~(\ref{eq:W_kC}), the lower bound of the spectral integral region, $k_{\rm{C}}$, is directly connected to the largest eddy size of turbulent motions as
\begin{equation}
	\ell_{\rm{C}} = 2\pi / k_{\rm{C}},
	\label{ell_C_def}
\end{equation}
which dominantly contributes to and determines the turbulent MHD energy and cross helicity. Strictly speaking, the correlation length of the cross helicity, $\ell_W$, can be different from that of the MHD energy, $\ell_K$. However, if we consider the fact that the characteristic lengths of turbulence represent the scales of the largest turbulent motion corresponding to the scales with largest magnitudes of the turbulent MHD-energy and cross-helicity spectral densities, we see that in order for the length scales of the MHD energy and cross helicity to be considerably different from each other, the spectral distributions of the MHD energy and cross helicity should be entirely different from each other. This is not the case, for instance, in the case of the solar-wind turbulence. Actually, in most cases of interests, these two length scale are similar to each other. In this sense, we can regard the characteristic lengths of the energy and cross helicity are approximately the same, and we denote $\ell_W \simeq \ell_K (\equiv \ell_{\rm{C}})$ hereafter. In this respect, extensive discussions of correlation length scales as well as time scales found in Matthaeus et al.\cite{mat1994} and Hossain et al.\cite{hos1995} are very important.

	With this point in mind, we perform the Lagrange derivatives and calculate the spectral integrals in Equation~(\ref{eq:W_kC}). Denoting the scaled wave number
\begin{equation}
	s = k / k_{\rm{C}},
	\label{eq:s_def}
\end{equation}
we have
\begin{eqnarray}
	\lefteqn{
	W 
	= 4 \cdot (2\pi)^{1/3} \sigma_{W0} 
	\varepsilon^{-1/3} \varepsilon_W k_{\rm{C}}^{2/3} 
	\int_{s \ge 1} \!\!\!ds\ s^{-5/3}
	}\nonumber\\
	& & - 4 \cdot 2 \pi \frac{\sigma_{W0}}
	{\omega_{\rm{sw}}\varepsilon_{\rm{sw}}^{1/3}} 
	k_{\rm{C}}^{7/3}
	\int_{s\ge 1}\!\!\! ds\ 
	s^{-7/3} \frac{D}{Dt} \left[ {
		\varepsilon^{-1/3}({\bf{x}};t) 
		\varepsilon_W({\bf{x}};t) 
		k_{\rm{C}}^{-11/3}
	} \right]
	\nonumber\\
	& & + 4 \cdot 2 \pi \frac{\sigma_{W0} \omega_{W0}}
	{(\omega_{\rm{sw}} \varepsilon_{\rm{sw}}^{1/3})^2}
	\varepsilon^{-1/3} \varepsilon_W k_{\rm{C}}^{-2}
		\int_{s \ge 1}\!\!\! ds\ s^{-7/3} 
	\frac{D}{Dt} \left[ {
	\varepsilon_W^{1/3}({\bf{x}};t) k_{\rm{C}}^{2/3}
	} \right].
	\label{eq:W_s}
\end{eqnarray}

	Using Equation~(\ref{ell_C_def}), we calculate Equation~(\ref{eq:W_s}) to obtain
\begin{eqnarray}
	\lefteqn{W
= 6 \cdot (2\pi)^{1/3} \sigma_{W0}
	\varepsilon^{-1/3} \varepsilon_W 
	\ell_{\rm{C}}^{2/3}
}\nonumber\\
& & + \frac{1}{(2 \pi)^{1/3}} 
	\frac{\sigma_{W0}}{\omega_{\rm{sw}} 
	\varepsilon_{\rm{sw}}^{1/3}}
	\varepsilon^{-1/3} \varepsilon_W 
	\ell_{\rm{C}}^{4/3}
	 \left\{ {
		\frac{1}{\varepsilon} \frac{D\varepsilon}{Dt}
	- \left[ {
	3 - A_W(\omega_{\rm{S}0}, \omega_{W0})
	} \right] \frac{1}{\varepsilon_W} 
		\frac{D\varepsilon_W}{Dt}
	} \right.
\nonumber\\
& & \left. {
	\hspace{40pt} + \left[ {11 
		- 2 A_W(\omega_{\rm{S}0}, \omega_{W0})
	} \right] \frac{1}{\ell_{\rm{C}}} 
		\frac{D\ell_{\rm{C}}}{Dt}
} \right\}.
	\label{eq:W_epsW_ellC}
\end{eqnarray}

	In hydrodynamic turbulence modeling, the turbulent energy $K_u (\equiv \langle {{\bf{u}}'{}^2} \rangle/2)$, its dissipation rate $\epsilon$, and the correlation or integral length $\ell_{\rm{C}}$ are equivalently important. For the purpose of closing the system of model equations, we can select any combination of $K_u$, $\epsilon$, and $\ell_{\rm{C}}$. This is known as the transferability of the model with respect to $K_u$, $\epsilon$, and $\ell_{\rm{C}}$. In order to satisfy this transferability requirement, these quantities should be connected with each other in an algebraic relation:
\begin{equation}
	K_u = C_K \epsilon^{2/3} \ell_{\rm{C}}^{2/3}.
	\label{eq:model_transf_HD}
\end{equation}
This transferability requirement gives a theoretical foundation of the energy-dissipation-rate equation (\ref{eq:eps_eq_hd}).\cite{yos1987}

	Relation (\ref{eq:model_transf_HD}) itself has been argued for a long time. As for the theoretical and experimental basis for this relation, the reader is referred to classical papers \cite{tay1938,kar1938} and also to Batchelor's book \cite{bat1953} and works cited therein. As for the simple physical arguments for this relation, see \cite{sre1995, sre1998, pea2004}. Here, we just point out that this relation is easily obtained if we assume the local equilibrium of turbulence; $P_{Ku} \simeq \epsilon$ with the notion of the mixing length. Then we can estimate the dissipation rate as
\begin{equation}
		\epsilon \simeq P_{Ku}
		= - \langle {u'{}^a u'{}^b} \rangle 
			\frac{\partial U^b}{\partial x^a}
		\sim uu \frac{u}{\ell}
	\label{eq:eps_estimate}
\end{equation}
($u$: characteristic intensity of turbulence, $\ell$: mixing length). This is equivalent to Equation~(\ref{eq:model_transf_HD}).

	In the context of theoretical derivation of $\epsilon$ equation, it is the algebraic property of Equation~(\ref{eq:model_transf_HD}) that is much more important. If the relation is not algebraic, we have no transferability among any combination of $K_u$, $\epsilon$, and $\ell_{\rm{C}}$ at all.

	 We expand the transferability requirement to the model equation related to the cross helicity. For this purpose, we solve Equation~(\ref{eq:W_epsW_ellC}) concerning $\ell_{\rm{C}}$ in a perturbational manner. At the lowest-order, we have
\begin{equation}
	W = 6 (2\pi)^{1/3} \sigma_{W0}
	\varepsilon^{-1/3} \varepsilon_W \ell_{\rm{C}}^{2/3}
	\label{eq:W_algeb_ell-W}
\end{equation}
or
\begin{equation}
	\ell_{\rm{C}} 
	= 6^{-3/2} (2\pi)^{-1/2} \sigma_{W0}^{-3/2}
	\varepsilon^{1/2} \varepsilon_W^{-3/2} W^{3/2}.
	\label{eq:W_algeb_W-ell}
\end{equation}
The transferability of model requires an algebraic relation among $W$, $\varepsilon$, $\varepsilon_W$, and $\ell_{\rm{C}}$. Equations (\ref{eq:W_algeb_ell-W}) and (\ref{eq:W_algeb_W-ell}) correspond to such a relation.

	Using Equation~(\ref{eq:W_algeb_W-ell}), we change expression (\ref{eq:W_epsW_ellC}) based on $\varepsilon$, $\varepsilon_W$, and $\ell_{\rm{C}}$ into the one based on $\varepsilon$, $\varepsilon_W$, and $W$. We require Equation~(\ref{eq:W_algeb_ell-W}) or (\ref{eq:W_algeb_W-ell}). As a result, we have
\begin{equation}
	\frac{D\varepsilon_W}{Dt}
	= C_{1} (\omega_{\rm{S}0}, \omega_{W0})
	\frac{\varepsilon_W}{\varepsilon} \frac{D\varepsilon}{Dt}
	+ C_{2}(\omega_{\rm{S}0}, \omega_{W0})
	\frac{\varepsilon_W}{W} \frac{DW}{Dt}
	\label{eq:eps_W_eq_C1C2}
\end{equation}
with
\begin{equation}
	C_{1}(\omega_{\rm{S}0}, \omega_{W0})
	= \frac{13 - A_W(\omega_{\rm{S}0}, \omega_{W0})}
	{39 - 8 A_W(\omega_{\rm{S}0}, \omega_{W0})},\;\;
	C_{2}(\omega_{\rm{S}0}, \omega_{W0})
	= \frac{33 - 6 A_W(\omega_{\rm{S}0}, \omega_{W0})}
	{39 - 8 A_W(\omega_{\rm{S}0}, \omega_{W0})}.
	\label{eq:C1C2_def}
\end{equation}
If the cross-helicity time scale can be treated as similar to the time scale associated with the Green's function:
\begin{equation}
	\omega_{\rm{S}0} \varepsilon^{1/3} 
	\simeq \omega_{W0} \varepsilon_W^{1/3},
	\label{eq:eps_sim_epsW}
\end{equation}
$A_W(\omega_{\rm{S}0}, \omega_{W0})$ [Equation~(\ref{eq:A_epsepsW_def})] can be estimated as
\begin{equation}
	A_W(\omega_{\rm{S}0}, \omega_{W0}) \simeq 1/2.
	\label{eq:A_sim_1/2}
\end{equation}
Then we have model coefficients
\begin{equation}
	C_{1}(\omega_{\rm{S}0}, \omega_{W0}) \simeq \frac{12}{35} \simeq 0.34,\;\;
	C_{2}(\omega_{\rm{S}0}, \omega_{W0}) \simeq \frac{6}{7} \simeq 0.86.
	\label{eq:C1C2_1/2}
\end{equation}

	If we substitute equations of $\varepsilon$ and $W$ into Equation~(\ref{eq:eps_W_eq_C1C2}), we finally obtain
\begin{equation}
	\frac{D\varepsilon_W}{Dt}
	= C_{\varepsilon_W 1} \frac{\varepsilon_W}{K} P_K
	- C_{\varepsilon_W 2} \frac{\varepsilon_W}{K} \varepsilon
	+ C_{\varepsilon_W 3} \frac{\varepsilon_W}{W} P_W
	- C_{\varepsilon_W 4} \frac{\varepsilon_W}{W} \varepsilon_W
	\label{eq:eps_W_eq_final}
\end{equation}
with
\begin{equation}
	C_{\varepsilon_W 1} = C_1 C_{\varepsilon 1},\;
	C_{\varepsilon_W 2} = C_1 C_{\varepsilon 1},\;
	C_{\varepsilon_W 3} = C_{\varepsilon_W 4} = C_2
	\label{eq:CepsW1CepsW2CepsW3}
\end{equation}
[$C_{\varepsilon 1}$ and $C_{\varepsilon 2}$ are given as Equation~(\ref{eq:eps_model_const})].

	From Equation~(\ref{eq:W_algeb_ell-W}), we shortly have
\begin{equation}
	\varepsilon_W
	= 6^{-1} (2\pi)^{-1/3} \sigma_{W0}^{-1} \frac{W}{\tau}
	\label{eq:eps_W_algeb_model}
\end{equation}
with
\begin{equation}
	\tau = \varepsilon^{-1/3} \ell_{\rm{C}}^{2/3}.
	\label{eq:eps_W_algeb_time}
\end{equation}
This corresponds to the simplest possible algebraic model [Equation~(\ref{eq:algebraic_eps_w})] for the cross-helicity dissipation rate. In other word, the simplest possible algebraic model for the cross-helicity dissipation corresponds to the scaling of Equation~(\ref{eq:sigmaW}).

	In the presence of a mean magnetic field, turbulence deviates from the isotropic state and the resulting anisotropic effects should be taken into account (See e.g., \cite{she1983}). As is shown in Appendix~\ref{sec:appendixC}, in the present formalism, we can take the effects of mean magnetic field and large-scale rotation into consideration, and obtain the the cross-helicity dissipation equation with such effects incorporated. Related to this point, we should make the following remarks. In this work, we assume an isotropic form Equation~(\ref{eq:sigmaW}) for the basic or lowest-order fields, ${\bf{u}}_{\rm{B}}'$ and ${\bf{b}}_{\rm{B}}'$, but the deviations from homogeneous isotropic state are taken into consideration from the higher-order contributions related to the inhomogeneity of mean velocity and the presence of the mean magnetic and vorticity fields. In this sense, the turbulence fields themselves are neither isotropic nor homogeneous in this work. However, this point never denies the importance of incorporating the anisotropy effects into the basic fields. This is a very interesting topic in the theoretical formulation of inhomogeneous turbulence, and will be reported in the forthcoming paper.

\section{A model for MHD turbulence}\label{sec:level5}
	One of the prominent features of turbulence is its wide ranges of scales. We
have continuous scales of motions ranging from the energy-containing scale, in
which energy is injected to turbulence through the inhomogeneity of mean fields,
to the dissipation or Kolmogorov scale where the dissipation plays a dominant
role. Owing to this breadth of scales, it is
impossible in the foreseeable future to simultaneously solve all the scales of
turbulence at the high Reynolds number encountered in the real-world flow of
interests. In such a situation, the notion of turbulence modeling provides a
useful tool for analyzing the real-world turbulence phenomena. In the turbulence
modeling, small-scale motions are modeled, and turbulence effects are
incorporated into the analysis of the large-scale or mean motions.

	In an MHD fluid, the mean fields obey

\noindent
Equation of continuity:
\begin{equation}
		\frac{\partial \bar{\rho}}{\partial t}
		+ \nabla \cdot \left( {	\bar{\rho} {\bf{U}}} \right)
		= 0,
	\label{eq:mean_continuity_eq}
\end{equation}
Momentum equation:
\begin{equation}
		\frac{\partial {\bf{U}}}{\partial t}
		+ \nabla \cdot \left( {
			{\bf{U}} {\bf{U}} - {\bf{B}} {\bf{B}}
		} \right)
		= - \nabla P_{\rm{M}}
		- \nabla \cdot \mbox{\boldmath$\cal{R}$}
		+ \nu \nabla^2 {\bf{U}},
	\label{eq:mean_vel_eq}
\end{equation}
Magnetic induction equation:
\begin{equation}
		\frac{\partial {\bf{B}}}{\partial t}
		= \nabla \times \left( {{\bf{U}} \times {\bf{B}}} \right)
		+ \nabla \times {\bf{E}}_{\rm{M}}
		+ \lambda \nabla^2 {\bf{B}},
	\label{eq:mean_mag_eq}
\end{equation}
Solenoidal condition for the magnetic field:
\begin{equation}
		\nabla \cdot {\bf{B}}
	= - \frac{1}{2} \left( {{\bf{B}} \cdot \nabla} \right) 
	\ln\bar{\rho}.
	\label{eq:mean_mag_sol}
\end{equation}
Here, $\bar{\rho}$ is the mean density. For the sake of simplicity, the
fluctuation part of the density has been neglected ($\rho'=0$). Measured in the Alfv\'{e}n-speed unit [Equation~(\ref{eq:alfven_unit})], the solenoidal condition of the 
original mean magnetic field:
\begin{equation}
		\nabla \cdot {\bf{B}}_\ast = 0
	\label{eq:solenoidal_B_ast}
\end{equation}
is expressed as Equation~(\ref{eq:mean_mag_sol}).

	Equations~(\ref{eq:mean_vel_eq}) and (\ref{eq:mean_mag_eq}) show that the
Reynolds stress $\mbox{\boldmath$\cal{R}$}$ and the turbulent electromotive force
${\bf{E}}_{\rm{M}}$ defined by Equations~(\ref{eq:Reynolds_strss_def}) and
(\ref{eq:emf_def}), represent the turbulence effects in the mean-field equations.
Estimating these correlations is of central importance in the study of
inhomogeneous turbulence. In order to close the system of equations, they should
be modeled in terms of the mean-field quantities. Here, we adopt
\begin{equation}
		{\cal{R}}^{\alpha\beta} 
		= \frac{2}{3} K_{\rm{R}} \delta^{\alpha\beta}
		- \nu_{\rm{K}} {\cal{S}}^{\alpha\beta}
		+ \nu_{\rm{M}} {\cal{M}}^{\alpha\beta},
	\label{eq:Reynolds_strss_exp}
\end{equation}
\begin{equation}
		{\bf{E}}_{\rm{M}} 
		= \alpha {\bf{B}}
		- \beta \nabla \times {\bf{B}}
		+ \gamma \nabla \times {\bf{U}},
	\label{eq:emf_exp}
\end{equation}
where $\mbox{\boldmath$\cal{S}$}$ and $\mbox{\boldmath$\cal{M}$}$ are the strain
rates of the mean velocity and the mean magnetic field:
\begin{subequations}
\begin{equation}
	{\cal{S}}^{\alpha\beta}
	= \frac{\partial U^a}{\partial x^b} 
	+ \frac{\partial U^b}{\partial x^a} 
	- \frac{2}{3} \delta^{\alpha\beta} \nabla \cdot {\bf{U}},
\end{equation}
\begin{equation}
	{\cal{M}}^{\alpha\beta}
	= \frac{\partial B^a}{\partial x^b} 
	+ \frac{\partial B^b}{\partial x^a}
	- \frac{2}{3} \delta^{\alpha\beta} \nabla \cdot {\bf{B}}.
\end{equation}
\end{subequations}

	We should note that the transport coefficients appearing in
Equations~(\ref{eq:Reynolds_strss_exp}) and (\ref{eq:emf_exp}), $\nu_{\rm{K}}$,
$\nu_{\rm{M}}$, $\alpha$, $\beta$, and $\gamma$, are not adjustable parameters,
but should be determined self-consistently through the dynamic properties of
turbulence. They in general depend on location and time. In this work,
we adopt model expressions for them as
\begin{equation}
	\alpha = C_\alpha \tau H,\;\;\;
	\beta = \frac{5}{7} \nu_{\rm{K}} = C_\beta \tau K,\;\;\;
	\gamma = \frac{5}{7} \nu_{\rm{M}} = C_\gamma \tau W,
	\label{eq:alpha_beta_gamma}
\end{equation}
where 
\begin{equation}
	H \equiv \langle {
		- {\bf{u}}' \cdot \mbox{\boldmath$\omega$}' 
		+ {\bf{b}}' \cdot {\bf{j}}'
	} \rangle,
	\label{eq:res_hel_def}
\end{equation}
and $\tau$ denotes the time scale of turbulence. The model constants
$C_\alpha$, $C_\beta$, and $C_\gamma$ are estimated as\cite{yos1998}
\begin{equation}
	C_\alpha \simeq 0.02,\;\;
	C_\beta \simeq 0.05,\;\;
	C_\gamma \simeq 0.04.
\end{equation}

	In the Reynolds-averaged turbulence model, properties of turbulence are
represented by some statistical quantities. We adopt the turbulent MHD energy
$K$, its dissipation rate $\varepsilon$, the turbulent cross helicity $W$, the cross-helicity dissipation rate, $\varepsilon_W$, and
the turbulent residual energy $K_{\rm{R}}$ as the turbulence statistical
quantities. The definition of them are given by $K$ [Equation~(\ref{eq:K_def})], $W$ [Equation~(\ref{eq:W_def})], 
\begin{equation}
		K_{\rm{R}} \equiv \frac{1}{2} \left\langle {
			{\bf{u}}'{}^2 - {\bf{b}}'{}^2
		} \right\rangle,
	\label{eq:K_R_def}
\end{equation}
and the dissipation rates of the turbulent MHD energy and cross helicity, $\varepsilon$ [Equation~(\ref{eq:eps_K_def})] and $\varepsilon_W$ [Equation~(\ref{eq:eps_W_def})], respectively.

	In this work, we assume the $\alpha$- or turbulent residual-helicity-related term in Equation~(\ref{eq:emf_exp}) is much smaller than the $\gamma$- or cross-helicity-related term, and neglect the former. So, we do not include the evolution equation of $H$. As for the modeling of $H$ equation and inclusion of it into the turbulence model, see \cite{yos1996,yos1998,yok2008}. Related to this point, the equations of kinetic helicity in HD turbulence and in MHD turbulence are an interesting subject. For these topics, the reader is referred to \cite{yok1993} for the HD case and \cite{oug2000} for the MHD case.

	If we take the effects of mean-density variation into account, the
evolution equations for the statistical quantities are written as
\begin{eqnarray}
		\lefteqn{
		\frac{\partial K}{\partial t}
		= - \left( { {\bf{U}} \cdot \nabla } \right) K
		- \frac{1}{6} \left( { 3K + K_{\rm{R}} } \right) 
			\nabla \cdot {\bf{U}}
		}\nonumber\\
		& & \hspace{10pt} + \frac{1}{2} ( {
			\nu_{\rm{K}} \mbox{\boldmath$\cal{S}$}^2 
			- \nu_{\rm{M}} \mbox{\boldmath$\cal{S}$}
				:\mbox{\boldmath$\cal{M}$}
		} )
		+ \beta {\bf{J}}^2 
		- \gamma \mbox{\boldmath$\Omega$} \cdot {\bf{J}}
		\nonumber\\
		& & \hspace{10pt} - \varepsilon
		+ \nabla \cdot \left( {
			W {\bf{B}}
		} \right)
		+ \nabla \cdot \left( {
			\nu_{\rm{K}} \nabla K
		} \right),
	\label{eq:K_eq_comp}
\end{eqnarray}
\begin{eqnarray}
		\lefteqn{
		\frac{\partial W}{\partial t}
		= - \left( { {\bf{U}} \cdot \nabla} \right) W
		- \frac{1}{2} W \nabla \cdot {\bf{U}}
		}\nonumber\\
		& & \hspace{10pt} + \frac{1}{2} \left( {
			\nu_{\rm{K}} \mbox{\boldmath$\cal{S}$}
			:\mbox{\boldmath$\cal{M}$} 
			- \nu_{\rm{M}} \mbox{\boldmath$\cal{M}$}^2 
		} \right)
		+ \beta {\bf{\Omega}}\cdot{\bf{J}}
		- \gamma {\bf{\Omega}}^2
		\nonumber\\
		& & \hspace{10pt} - \varepsilon_W
		+ \nabla \cdot \left( { K {\bf{B}} } \right)
		+ \nabla \cdot \left( { 
			\frac{\nu_K}{\sigma_{W}} \nabla W 
		} \right),
	\label{eq:W_eq_comp}
\end{eqnarray}
\begin{eqnarray}
		\lefteqn{
		\frac{\partial K_{\rm{R}}}{\partial t}
		= - \left( {\bf{U} \cdot \nabla} \right) K_{\rm{R}}
		- \frac{1}{6} \left( {K + 3 K_{\rm{R}}} \right) 
			\nabla \cdot {\bf{U}}
		}\nonumber\\
		& & \hspace{10pt} + \frac{1}{2} \nu_{\rm{R}} ( {
			\mbox{\boldmath$\cal{S}$}^2 
			- \mbox{\boldmath$\cal{M}$}^2 
		} )
		- \varepsilon_{\rm{R}}
		+ \nabla \cdot \left( {
			\frac{\nu_{\rm{K}}}{\sigma_{\rm{R}}} \nabla K_{\rm{R}}
		} \right),
	\label{eq:KR_eq_comp}
\end{eqnarray}
where $\mbox{\boldmath$\cal{S}$}^2 = {\cal{S}}^{ab} {\cal{S}}^{ab}$, 
$\mbox{\boldmath$\cal{S}$}:\mbox{\boldmath$\cal{M}$} = {\cal{S}}^{ab}
{\cal{M}}^{ab}$, etc. Here
$\nu_{\rm{R}}$ denotes the residual turbulent viscosity defined by
\begin{equation}
	\nu_{\rm{R}} = \nu_{\rm{K}} \frac{K_{\rm{R}}}{K}.
\end{equation}
We should note that Equations~(\ref{eq:K_eq_comp}) and (\ref{eq:W_eq_comp}) are very similar to equations found in Matthaeus et al.\cite{mat1994}

	The dissipation rate of the turbulent energy, $\varepsilon$, in Equation~(\ref{eq:K_eq_comp}) is given by Equation~(\ref{eq:eps_eq_mhd}). The dissipation rate of the turbulent cross helicity, $\varepsilon_W$
[Equation~(\ref{eq:eps_W_def})], is estimated by Equation~(\ref{eq:algebraic_eps_w}) or
alternatively, by Equation~(\ref{eq:eps_W_eq_final}).

	The dissipation rate of $K_{\rm{R}}$, $\varepsilon_{\rm{R}}$, in Equation~(\ref{eq:KR_eq_comp}) can be expressed as\cite{yok2006}
\begin{equation}
	\varepsilon_{\rm{R}}
	= \left( {
		1 + C_{r} \frac{{\bf{B}}^2}{K}
	} \right) \frac{\varepsilon}{K} K_{\rm{R}}.
	\label{eq:eps_R_exp}
\end{equation}
Here, $C_{r}$ and $\sigma_{\rm{R}}$ are positive model constants.
The large-scale behavior of $K_{\rm{R}}$ depends on the choice of these
constants. At this stage of modeling, we do not insist on fine tuning of
the model constants, but roughly put them 
\begin{equation}
	C_{r} = 0.01,\;\; \sigma_{\rm{R}} = 1.0.
	\label{eq:res_en_model_const}
\end{equation}

	Equation~(\ref{eq:eps_R_exp}) as a whole will destroy the turbulent residual
energy, and return MHD turbulence to equipartition. The first part
in the parentheses of Equation~(\ref{eq:eps_R_exp}) represents the $K_{\rm{R}}$
destruction due to the eddy distortion. The second part represents the
destruction of $K_{\rm{R}}$ due to the mean magnetic field. The
latter corresponds to the Alfv\'{e}n effect, in which the presence of the mean
magnetic field leads MHD turbulence to an equipartition state between the
kinetic and magnetic energies.\cite{iro1964,kra1965,pou1976,yok2006} In connection with the anisotropy and residual energy, several numerical studies have recently been performed in the presence of a uniform magnetic field. For this interesting topic, see \cite{mue2003,big2008a,big2008b} and works cited therein.

	Equation~(\ref{eq:eps_R_exp}) may be reinterpreted as the modulation of
MHD-turbulence time scale due to the mean magnetic field:
\begin{equation}
		\tau = \frac{K}{\varepsilon}\;\; \to\;\;
		\left( {
			1 + \chi \frac{{\bf{B}}^2}{K}
		} \right)^{-1} \frac{K}{\varepsilon},
	\label{eq:modulation_time}
\end{equation}
where $\chi (= C_r)$ represents the synthesization ratio of time scales.
Equation~(\ref{eq:res_en_model_const}) infers that we should put the ratio as
\begin{equation}
	\chi = O(10^{-2}).
\end{equation}
As for an example considering the synthesized time scale of MHD turbulence,
you are referred to \cite{yok2008} and references cited therein.

\section{An application to the solar-wind turbulence}\label{sec:level6}

\subsection{\it\textbf{Solar-wind turbulence}}\label{sec:level6-1}
\begin{figure}[htb]
\centering
\includegraphics[width=.75\textwidth]{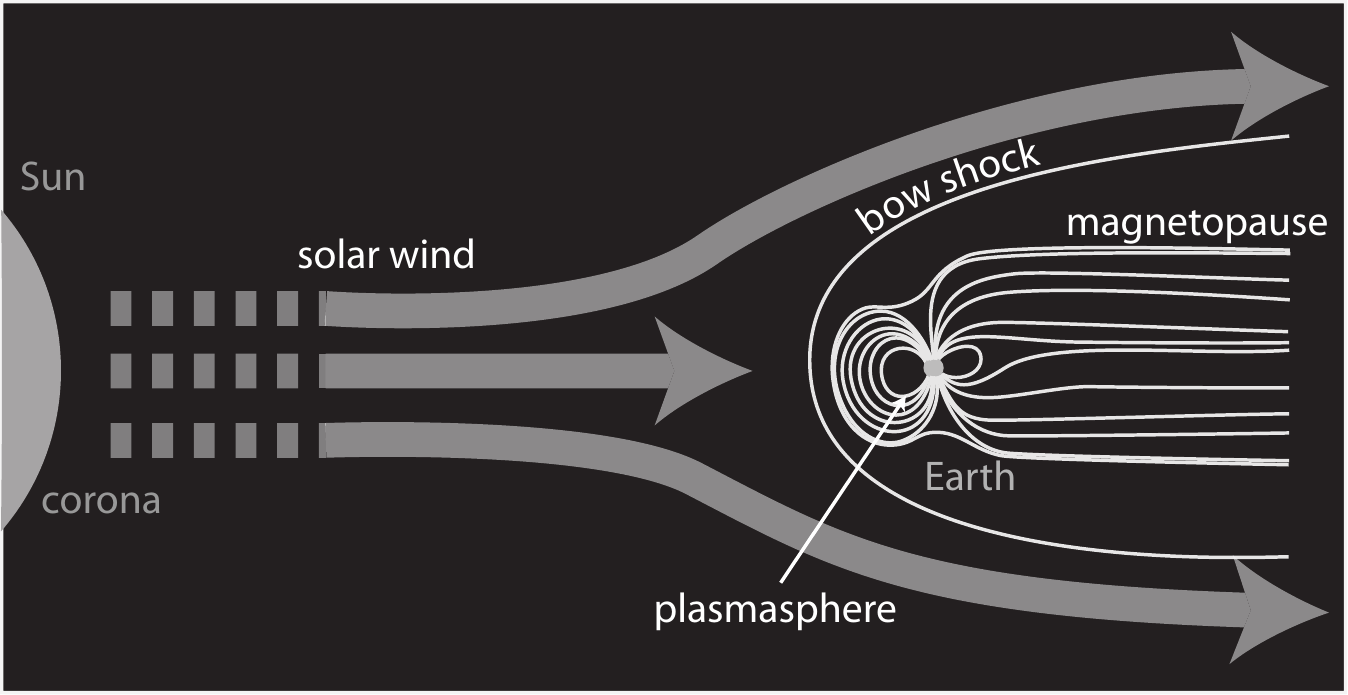}
\caption{Solar wind and terrestrial environments. The solar wind, a high speed
plasma flow of several hundreds ${\rm{km\ s}^{-1}}$, interacts with the
terrestrial magnetosphere to induce several phenomena such as substorms, auroras.}
	\label{fig:sun_earth}
\end{figure}

	Solar wind is a continuous plasma flow blown away from the coronal bases to the
solar-system space. Its origin is considered to be the violent magnetic
activities on the solar surface. The streams from the low- or mid-latitude region
whose typical speed is $400\ {\rm{km\ s}^{-1}}$ are called the slow wind. On the
other hand, the streams from the high-latitude region such as coronal hole whose
typical speed is $800\ {\rm{km\ s}^{-1}}$ are called the fast wind. It is known
that the slow wind has a large velocity shear, while the fast wind has
substantially no velocity shears. The influence of an explosive magnetic activity
on the solar surface is conveyed to the magnetosphere of the Earth within several
days, and terrestrial environments may be affected much. A direct exposure to a
high-energy particle convected by the solar wind is quite harmful to astronauts
and electric devices on the satellite. So it is highly desirable to predict the
behavior of the solar wind: A problem of the space weather
(Figure~\ref{fig:sun_earth}).

	Satellite observations have revealed the statistical properties of the
solar-wind turbulence including the velocity, magnetic field, density, etc. It has
been investigated how MHD turbulence evolves in the large-scale velocity and
magnetic-field structures.\cite{bel1971,rob1987,tum1995} According to the
observations, the solar-wind turbulence shows a strong Alfv\'{e}nicity near the
Sun. Namely, there is a strong correlation between the velocity and
magnetic-field fluctuations, and equipartition between the kinetic and magnetic
turbulent energies is realized. At the same time, it is pointed out that this
Alfv\'{e}nicity decays as the heliocentric distance increases.\cite{tum1995} One
of the main unsolved problems in this field is the spatial evolutions of the
turbulent cross helicity $W$ and the Alfv\'{e}n ratio $r_{\rm{A}}$.\cite{rob1990}

	The magnitude of the scaled turbulent cross helicity, $|W|/K$, whose value is
almost unity near the Sun, decreases as the heliocentric distance increases. In
the region with larger mean-velocity shear, the decay rate of $|W|/K$ is larger.
The value of $|W|/K$ far from the Sun is small ($0 - 0.2$). On the other hand,
in the region with smaller mean-velocity shear, the decay of $|W|/K$ is
suppressed. The value of $|W|/K$ remains to be large ($0.4 - 0.7$) even far from
the Sun.\cite{golbe1995} The previous models could not
reproduce this large value of the scaled $W$ in the region with small or almost no
mean-velocity shear.

	On the other hand, the Alfv\'{e}n ratio $r_{\rm{A}}$ is defined by the ratio
of the turbulent kinetic and magnetic energies:
\begin{equation}
	r_{\rm{A}} 
	\equiv \frac{\langle {\bf{u}}'{}^2 \rangle} {\langle {\bf{b}}'{}^2 \rangle}.
\end{equation}
This ratio is almost unity near the Sun, then decreases with the heliocentric
distance. At about $3\ {\rm{AU}}$ (astronomical unit: $1\ {\rm{AU}} = 1.5 \times
10^{11}\ {\rm{m}}$) from the Sun, $r_{\rm{A}}$ reaches $0.5$ and thereafter
remains to be the same ($\sim 0.5$) with the heliocentric distance as long as the
observations exist:
\begin{equation}
		r_{\rm{A}} \simeq 0.5\;\;\; 
		\mbox{for}\;\;\; 
		r \gtrsim 3\; {\rm{AU}}.
	\label{eq:alfven_ratio_1over2}
\end{equation}
Previous research could not properly explain this stationary value of
$0.5$.\cite{tum1995}

	In this work, we address these problems from the viewpoint of the turbulence
model. In particular, we examine how the large-scale spatial evolution varies as
the model expression for the turbulent cross-helicity dissipation rate changes.

\subsection{\it\textbf{Numerical simulations}}\label{sec:level6-2}
	We examine the evolution of the turbulent statistical quantities
under the prescribed mean velocity and magnetic field. The basic set-up of the numerical simulation is the same as the previous work, so, for the details of the numerical simulation the reader is referred to \cite{yok2007,yok2008}. For the choice of the mean fields, we fully utilize the current theory and modeling of the solar wind. The magnetic rotator model of the solar wind takes the effects of rotation and
magnetic field into consideration, and known to be a good
approximation for the mean velocity and magnetic field of the solar
wind.\cite{par1958,web1967} However, its large-scale velocity and magnetic-field
shears are weak as compared with the large-scale shears in the slow--fast-wind
and magnetic-sector boundaries. So, the adopted mean-field profile is suitable
for representing relatively calm flow fields within one magnetic sector.

	First, we adopt the algebraic expression [Equation~(\ref{eq:algebraic_eps_w})] for
the dissipation of the turbulent cross helicity. In this case,
Equation~(\ref{eq:W_K_constraint_basic}) is rewritten as
\begin{equation}
		C_W > 1.
	\label{eq:C_w_condition_alg}
\end{equation}
Namely, the model constant $C_W$ for $\varepsilon_W$ in the $W$ equation should be
larger than the counterpart ($= 1$) for $\varepsilon$ in the $K$ equation.

	In this work, we adopt
\begin{equation}
		C_W = 1.4,\;\; \sigma_W = 1.0.
	\label{eq:C_W_sigma_W_consts}
\end{equation}
Here, we should note the following point. In the previous work, $C_W
= 1.8$ was adopted for the algebraic model constant for
$\varepsilon_W$.\cite{yok2007} As was mentioned above, the assumed mean fields
correspond to a weak-shear case in which $W$ is expected to show no considerable
decay. If we consider this fact, we see that the less dissipative value for $C_W$
[Equation~(\ref{eq:C_W_sigma_W_consts})] is more suitable than the more dissipative
value of $C_W = 1.8$ previously adopted.

\subsection{Results}\label{sec:level6-3}
\subsubsection{Alfv\'{e}n ratio}\label{sec:level6-3-1}
\begin{figure}[htb]
\centering
\includegraphics[width=.50\textwidth]{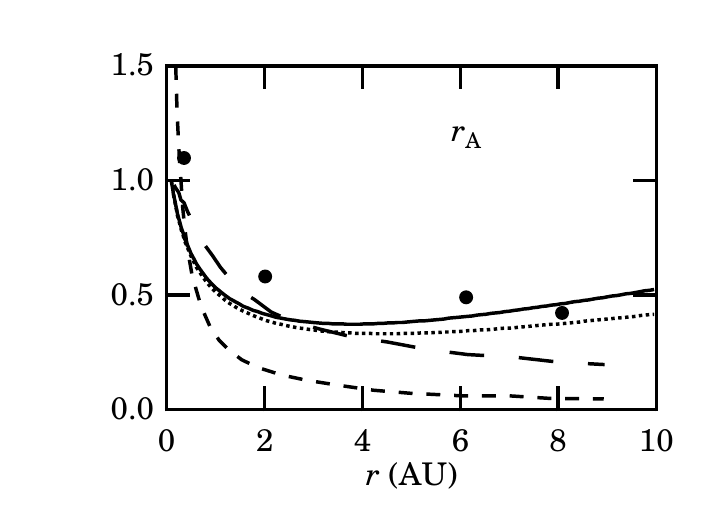}
\caption{The radial evolution of the Alfv\'{e}n ratio $r_{\rm{A}}$.
$\bullet$: observations by Roberts et al.\cite{rob1987}; ------: the present
model; $\cdots\cdots$: the present model with enhanced shear. Previous work, - -
-: Zhou and Matthaeus\cite{zho1990}; ---\hspace{8pt}---\hspace{8pt}---: Tu and
Marsch\cite{tum1993}. For much recent work for comparison, the reader is referred to Breech et al.\cite{bre2008}}
	\label{fig:alfven_ratio_result}
\end{figure}

The simulated Alfv\'{e}n ratio $r_{\rm{A}}$ is shown in Figure~\ref{fig:alfven_ratio_result} with the comparisons with the observations and with the previous work.

	The evolution of Alfv\'{e}n ratio $r_{\rm{A}}$ does not depend so much on the value of $C_W$. In the case of higher velocity shear (the dotted line in Figure~\ref{fig:alfven_ratio_result}), the large scale evolution of $r_{\rm{A}}$ shows a more stationary behavior in space, which is in better agreement with the observations. As was already mentioned, if we made the fine tuning of $C_r$ value, the agreement of the simulation results and observations becomes better. In this work, however, we do not do such fine tuning, and made only a rough estimate of the constants as in Equation~(\ref{eq:res_en_model_const}).

\begin{figure}[htb]
\centering
\includegraphics[width=.60\textwidth]{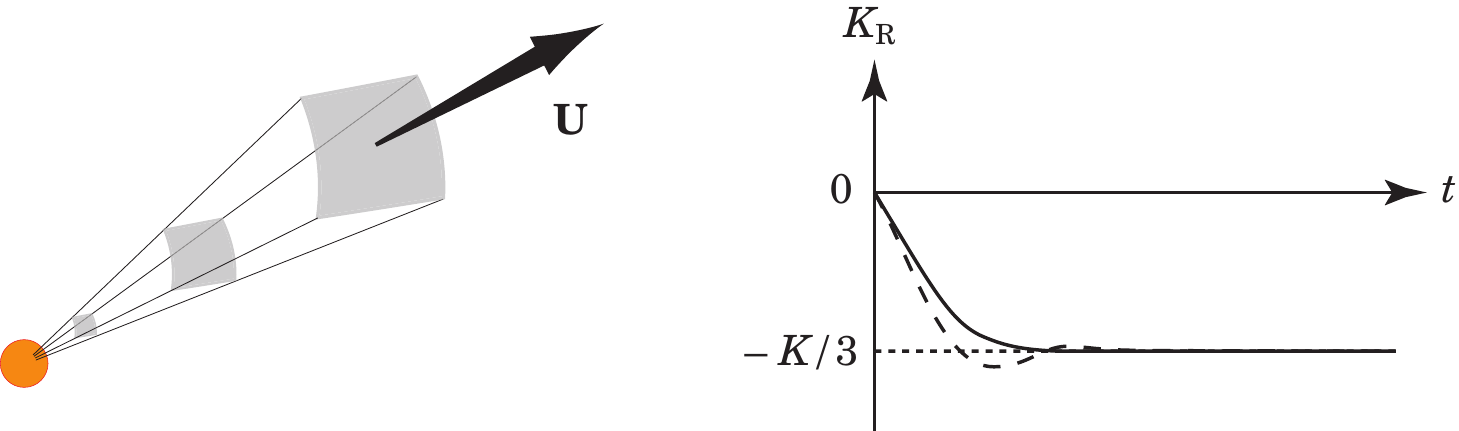}
\caption{Expansion effect and equilibrium of the turbulent residual energy.
(Left) Flow expansion. (Right) Schematically depicted equilibrium state of the turbulent residual energy $K_{\rm{R}}$. From the equipartition state ($K_{\rm{R}} = 0$) near the Sun, $K_{\rm{R}}$ starts decreasing towards the equilibrium value of $K_{\rm{R}} = -K/3$.}
	\label{fig:expansion_effect}
\end{figure}

	In Equation~(\ref{eq:KR_eq_comp}), the turbulent residual energy equation contains a term related to the expansion of the large-scale flow through $\nabla \cdot {\bf{U}}$: 
\begin{equation}
		P_{K\rm{R1}} = - \frac{1}{6} (K + 3 K_{\rm{R}}) \nabla \cdot {\bf{U}}.
	\label{eq:P_KR1_def}
\end{equation}
In the solar-wind turbulence, the dominance of Alfv\'{e}n-wave effects would give an equipartition of the kinetic and magnetic energies near the Sun. Therefore, we have $K_{\rm{R}} = 0$ for the inner boundary near the Sun. Since the turbulent MHD energy is always positive ($K > 0$), we have negative contribution from $P_{K\rm{R1}}$ near the Sun:
\begin{equation}
		P_{K\rm{R1}} 
		\simeq - \frac{1}{6} K \nabla \cdot {\bf{U}} < 0,
	\label{eq:P_KR1_near}
\end{equation}
which produces a negative residual energy ($K_{\rm{R}} < 0$), i.e., magnetic dominance there. Due to this negative production, the magnitude of negative $K_{\rm{R}}$ increases until its value reaches
\begin{equation}
		K + 3 K_{\rm{R}} = 0.
	\label{eq:K_R_equilibrium}
\end{equation}
If the magnitude of $K_{\rm{R}}$ becomes larger than $K/3$, the sign of the production $P_{K\rm{R1}}$ reverses to positive, then the production of negative $K_{\rm{R}}$ is stopped. This infers that $K_{\rm{R}} = - K/3$ is an equilibrium state of MHD turbulence in the expanding solar-wind geometry with equipartition at the inner boundary (Figure~\ref{fig:expansion_effect}).

	We should note the fact that the Alfv\'{e}n ratio of $0.5$ corresponds to the $K_{\rm{R}}/K$ of $-1/3$ as
\begin{equation}
		r_{\rm{A}} 
		= \frac{\langle {{\bf{u}}'{}^2} \rangle}
			{\langle {{\bf{b}}'{}^2} \rangle} 
		= \frac{1}{2}\;\;
		\leftrightarrow\;\;
		\frac{K_{\rm{R}}}{K}
		= \frac{\langle {{\bf{u}}'{}^2 - {\bf{b}}'{}^2} \rangle}
			{\langle {{\bf{b}}'{}^2 + {\bf{u}}'{}^2} \rangle}
		= \frac{1-2}{1+2}
		= - \frac{1}{3}.
	\label{eq:rA_KR_relation}
\end{equation}
Equation~(\ref{eq:K_R_equilibrium}) is suggestive for the explanation of the observed value of the Alfv\'{e}n ratio $r_{\rm{A}} \simeq 0.5$. In the expanding solar-wind flow, due to the negative $K_{\rm{R}}$ production by $P_{K\rm{R1}}$, a state of magnetic dominance is realized. However, the magnitude of $K_{\rm{R}}$ can not exceed $K/3$, since the negative production associated with a flow expansion vanishes when $K_{\rm{R}}$ reaches $-K/3$. Equation~(\ref{eq:K_R_equilibrium}) reflects an equilibrium state of MHD turbulence in this geometry. Once this state has been realized, the flow expansion or compressibility makes no contributions to the Alfv\'{e}n-ratio evolution, and the turbulence evolves according to the balance of the other evolution mechanisms intrinsic to incompressible MHD turbulence. In other words, the value of Alfv\'{e}n ratio is determined by the expansion effect with the inner boundary condition. However, once turbulence has fell into the state of $r_{\rm{A}} \simeq 0.5$, the compressibility effects disappear, and turbulence shows a state stationary in space according to the incompressible balances.

\subsubsection{Cross helicity}\label{sec:level6-3-2}
\begin{figure}[htb]
\centering
\includegraphics[width=.50\textwidth]{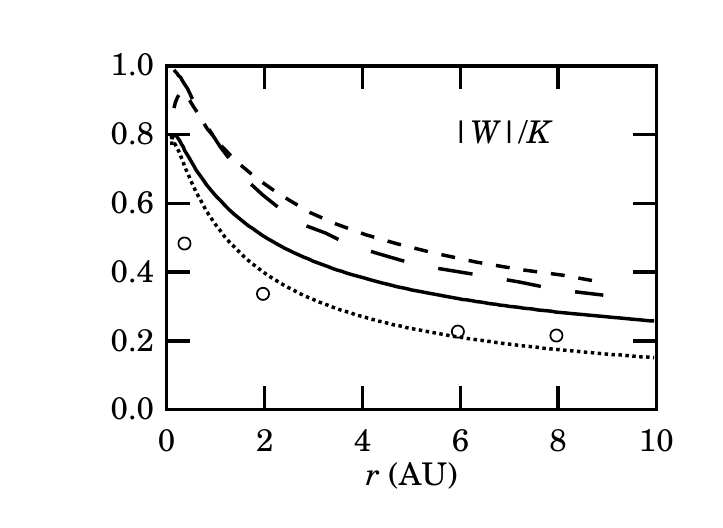}
\caption{The radial evolution of the scaled turbulent cross helicity $W$.
$\circ$: observations by Roberts et al.\cite{rob1987}; ------: the present
model; $\cdots\cdots$: the present model with enhanced shear. Previous work, - -
-: Zhou and Matthaeus\cite{zho1990}; ---\hspace{8pt}---\hspace{8pt}---: Tu and
Marsch\cite{tum1993}. For much recent work for comparison, the reader is referred to Breech et al.\cite{bre2008}}
	\label{fig:cross_helicity_result}
\end{figure}

			The simulated cross helicity $W$ is shown in
Figure~\ref{fig:cross_helicity_result} with the comparisons with the observations
and with the previous work.

	The radial evolution of the cross helicity in the higher velocity shear shows
better agreement with the observations. The mean fields adopted in this work
correspond to a weak-shear case, while the observations were performed in
situations with stronger shears. Then, the tendency that the simulation results
in the high-shear case are in better agreement is preferable. This agrees with
the observational findings that in the fast solar wind, which has a small velocity
shear, the scaled turbulent cross helicity $|W|/K$ is kept large without decaying. In the
simulation with a weak shear, $|W|/K$ is about 0.5 at 3 AU and about 0.3 at 10 AU, which are kept relatively large values.

	From this agreement of the numerical simulation with the observations, we see
that the algebraic-model expression for $\varepsilon_W$
[Equation~(\ref{eq:algebraic_eps_w})] is good enough, as far as the application to the
solar wind is concerned.

\subsection{Cross helicity and velocity shear}\label{sec:level6-4}
	As was referred to in Section~6.1, the magnitude of the scaled turbulent cross helicity $|W|/K$ decays much in the slow-wind region where the velocity shear is strong, while $|W|/K$ decays less in the fast-wind region where the shear is weak.\cite{golbe1995} This basic tendency can be elucidated from the viewpoint of turbulence model through a simple argument.

	The behavior of $|W|/K$ is governed by Equation~(\ref{eq:scaled_W_eq}). Among the terms, we pick up the production-related terms which are directly connected to the
mean-velocity shear as
\begin{equation}
	\frac{D}{Dt} \frac{W}{K}
	= \left( {\frac{1}{W} P_W - \frac{1}{K} P_K} \right)
		\frac{W}{K} 
	- \cdots.
	\label{eq:scaled_W_eq_prod}
\end{equation}
If we take the compressibility effect into consideration, the equation of $K$
and $W$ are expressed as Equations~(\ref{eq:K_eq_comp}) and (\ref{eq:W_eq_comp}). The production rates added by the compressibility- or $\nabla \cdot {\bf{U}}$-related
terms in Equations~(\ref{eq:K_eq_comp}) and (\ref{eq:W_eq_comp}) are given as
\begin{equation}
	P_K = - \frac{1}{6} \left( {
		3 K + K_{\rm{R}}
	} \right) \nabla \cdot {\bf{U}}
	+ \frac{7}{10} C_\beta \frac{K^2}{\varepsilon} {\mbox{\boldmath$\cal{S}$}}^2
	- \frac{7}{10} C_\gamma \frac{KW}{\varepsilon}
		\mbox{\boldmath$\cal{S}$}:\mbox{\boldmath$\cal{M}$},
	\label{eq:P_K_comp}
\end{equation}
\begin{equation}
	P_W = - \frac{1}{2} W \nabla \cdot {\bf{U}}
	+ \frac{7}{10} C_\beta \frac{K^2}{\varepsilon}
	\mbox{\boldmath$\cal{S}$}:\mbox{\boldmath$\cal{M}$}
	- \frac{7}{10} C_\gamma \frac{KW}{\varepsilon} \mbox{\boldmath$\cal{M}$}^2,
	\label{eq:P_W_comp}
\end{equation}
where use has been made of Equations~(\ref{eq:Reynolds_strss_exp}), (\ref{eq:emf_exp}),
and (\ref{eq:alpha_beta_gamma}) with time scale $\tau = K / \varepsilon$. We
substitute Equations~(\ref{eq:P_K_comp}) and (\ref{eq:P_W_comp}) into
Equation~(\ref{eq:scaled_W_eq_prod}) with the assumption that the mean velocity shear
is much larger than the mean magnetic shear:
\begin{equation}
		\mbox{\boldmath$\cal{S}$}^2 \gg
		|\mbox{\boldmath$\cal{S}$}:\mbox{\boldmath$\cal{M}$}|,\;
		\mbox{\boldmath$\cal{M}$}^2.
\end{equation}
Then, we have a contribution from the mean-velocity shear as
\begin{equation}
	\frac{D}{Dt} \frac{W}{K}
	= \left( {
		\frac{1}{6} \frac{K_{\rm{R}}}{K} \nabla \cdot {\bf{U}}
	- \frac{7}{10} C_\beta \frac{K}{\varepsilon} 
		\mbox{\boldmath$\cal{S}$}^2
		} \right) \frac{W}{K}
	+ \cdots.
	\label{eq:scld_W_shear}
\end{equation}
The first or $\nabla \cdot {\bf{U}}$-related term in the parentheses of
Equation~(\ref{eq:scld_W_shear}) represents the expansion effect. In an
expansion flow such as solar wind, $\nabla \cdot {\bf{U}} > 0$. The residual
energy $K_{\rm{R}}$ in the solar wind is nearly equal to zero near the Sun, and
$K_{\rm{R}} / K = -1/3$ far from the Sun [Equation~(\ref{eq:alfven_ratio_1over2})].
So, the contribution from the $\nabla \cdot {\bf{U}}$-related term is negative in
the solar wind case. Namely, the expanding flow coupled with $K_{\rm{R}} \le 0$
(the magnetic-energy dominance) leads to a decrease in the magnitude of $W/K$.
Equation~(\ref{eq:scld_W_shear}) can be rewritten as
\begin{eqnarray}
	\frac{D}{Dt} \frac{W}{K} = 
		\left\{ { \everymath{\displaystyle}
		\begin{array}{ll}
		- \frac{7}{10} C_\beta \frac{K}{\varepsilon} 
		\mbox{\boldmath$\cal{S}$}^2
		 \frac{W}{K} + \cdots,
		& (\mbox{near the Sun})\\
	 \rule{0ex}{5ex}- \left( {
		\frac{1}{18} \nabla \cdot {\bf{U}}
		+ \frac{7}{10} C_\beta \frac{K}{\varepsilon} 
		\mbox{\boldmath$\cal{S}$}^2
		} \right) \frac{W}{K} + \cdots.
		& (\mbox{far from the Sun})
		\end{array}
		} \right. 
\end{eqnarray}
 Since both
$C_\beta$ and 
$K/\varepsilon$ are positive, in the presence of the mean-velocity shear,
irrespective of the sign of $W/K$, it works for a decrease in the magnitude of $W/K$.

	If we scrutinize the large-scale velocity-shear effects on the production rate
of the turbulent cross helicity $W$, we see that whether $W$ may increase or decrease depends on the coupling of
the mean velocity and magnetic-field shears [the second or
$\mbox{\boldmath$\cal{S}$}:\mbox{\boldmath$\cal{M}$}$-related term in
Equation~(\ref{eq:P_W_comp})]. At the same time, the turbulent MHD energy
$K$ always increases in the presence of the mean-velocity shear
[the second or $\mbox{\boldmath${\cal{S}}$}^2$-related term in
Equation~(\ref{eq:P_K_comp})]. As this result, the magnitude of  the turbulent cross
helicity scaled by the turbulent MHD energy,
$|W|/K$, will decrease in the presence of the mean-velocity shear in the primary
sense. The present model reflects this mechanism properly. So, this model is
particularly promising for describing the evolution of the solar-wind turbulence.

\section{Conclusion}\label{sec:level7}
	The evolution of turbulent cross helicity ($W \equiv \langle {{\bf{u}}' \cdot {\bf{b}}'} \rangle$) was investigated from the viewpoint of generation and destruction mechanism of it. In particular, two possibilities of expressing the dissipation rate of $W$, $\varepsilon_W$, were presented; (i) the algebraic model and (ii) the evolution equation for $\varepsilon_W$. It was shown that both model expressions can be systematically derived from the zeroth- and first-order calculations of the statistical analytical theory of inhomogeneous turbulence, respectively. Validity of the model expressions was examined with the aid of a turbulence model constituted by four one-point turbulent statistical quantities (the turbulent MHD energy $K$, its dissipation rate $\varepsilon$, the turbulent cross helicity $W$, and the turbulent residual energy $K_{\rm{R}}$). It was shown that, as far as the application to the solar-wind turbulence is concerned, the algebraic model for $\varepsilon_W$ with one model constant gives results plausible enough. Dependence of the cross-helicity evolution on the large-scale velocity structure was also discussed. In the context of the solar wind, it was shown that both of flow expansion and large-scale velocity shear contribute to decreasing the magnitude of the scaled cross helicity, which confirmed earlier results.\cite{mat2004,bre2005,bre2008} More detailed expressions for the $\varepsilon_W$ equation were also indicated from the higher-order calculation of the statistical analytical theory of inhomogeneous turbulence.

\section*{Acknowledgments}
	The author would like to thank the organizing and executive
committees of the sixth international symposium on Turbulence and Shear Flow
Phenomena (TSFP-6) held at Seoul for providing him with a good opportunity to
present some part of this work at the well-organized symposium. His thanks are
also due to the guest editors for TSFP-6 focus issue of the Journal of Turbulence
(JoT), Rainer Friedrich and Arne Johansson, for invitimg him to submit the work
as a paper to the issue. The author would like to thank anonymous referees for their precious comments based on deep insight and wide knowledge on turbulence, which made the presentation of this paper much improved. Part of this work was performed during the periods the author stayed at the Nordic Institute for Theoretical Physics (NORDITA): One for a
NORDITA program on ``Solar and stellar dynamo and cycles" (Sep.\ 26 --
Oct.\ 26, 2009) and another as an invited researcher (Feb., 2010).


\appendix
\setcounter{equation}{0}\renewcommand\theequation{\thesection\arabic{equation}}%
\setcounter{section}{0}\renewcommand\thesection       {\Alph{section}}
\section{Mean magnetohydrodynamic (MHD) energy and cross-helicity equations}\label{sec:appendixA}
	The evolution equation of the turbulent cross helicity density $\langle {{\bf{u}}' \cdot {\bf{b}}'} \rangle (\equiv W)$ is written as Equation~(\ref{eq:transport_eq_G}) with Equation~(\ref{eq:W_eq_const}). Since the total amount of the cross helicity, $\int_V {\bf{u}} \cdot {\bf{b}} dV$, as well as that of the magnetohydrodynamic (MHD) energy, $\int_V ({\bf{u}}{}^2 + {\bf{b}}{}^2) / 2 dV$, is an inviscid invariant of the MHD equation, we can draw a clear picture for the $W$ evolution. The production of the turbulent cross helicity $\langle {{\bf{u}}' \cdot {\bf{b}}'} \rangle$ arises from a kind of cascading process in turbulence as the sink or drain of the mean cross helicity ${\bf{U}} \cdot {\bf{B}} (\equiv W_{\rm{M}})$. 

	This is an argument similar to the one for the evolution of the turbulent kinetic helicity density $\langle {{\bf{u}}' \cdot \mbox{\boldmath$\omega$}'} \rangle$ in the hydrodynamic (HD) turbulence. In the HD case, the total amount of the kinetic helicity $\int_V {\bf{u}} \cdot \mbox{\boldmath$\omega$} dV$, as well as that of the energy $\int_V {\bf{u}}{}^2 /2 dV$, is an inviscid invariant. In the HD case, such a clear picture was presented with the evolution equations of the turbulent and mean helicity densities, $\langle {\bf{u}}' \cdot \mbox{\boldmath$\omega$}' \rangle$ and ${\bf{U}} \cdot \mbox{\boldmath$\Omega$}$.\cite{yok1993}

	In order to see the evolution picture of the MHD energy and cross helicity clearly, here we write down the equations of the mean MHD energy and cross helicity densities:
\begin{equation}
		K_{\rm{M}} \equiv ({\bf{U}}^2 + {\bf{B}}^2)/2,
	\label{eq:mean_en_def}
\end{equation}
\begin{equation}
		W_{\rm{M}} \equiv {\bf{U}} \cdot {\bf{B}}.
	\label{eq:mean_cross_hel_def}
\end{equation}

	From mean velocity and magnetic field equations~(\ref{eq:mean_vel_eq}) and (\ref{eq:mean_mag_eq}), the evolution equations of the mean MHD energy and cross-helicity densities, $K_{\rm{M}}$ and $W_{\rm{M}}$, are obtained as
\begin{equation}
		\left( {
		\frac{\partial}{\partial t}
		+ {\bf{U}} \cdot \nabla
		} \right) G_{\rm{M}}
		= P_{G\rm{M}}
		- \varepsilon_{G\rm{M}}
		+ T_{G\rm{M}}
	\label{eq:}
\end{equation}
with $G_{\rm{M}} = (K_{\rm{M}}, W_{\rm{M}})$. Here, $P_{G\rm{M}}$, $\varepsilon_{G\rm{M}}$, and $T_{G\rm{M}}$ are the production, dissipation, and transport rates of the mean MHD energy and cross helicity, respectively. They are defined by
\begin{subequations}\label{eq:P_eps_T_GM}
\begin{equation}
		P_{K\rm{M}}
		= + {\cal{R}}^{ab} \frac{\partial U^b}{\partial x^a}
		+ {\bf{E}}_{\rm{M}} \cdot {\bf{J}}
		= - P_K,	
	\label{eq:product_KM}
\end{equation}
\begin{equation}
		\varepsilon_{K\rm{M}}
		= \nu \left( {\frac{\partial U^a}{\partial x^b}} \right)^2
		+ \lambda \left( {\frac{\partial B^a}{\partial x^b}} \right)^2,
	\label{eq:dissip_KM}
\end{equation}
\begin{equation}
		T_{K\rm{M}} = T_{K\rm{MT}} + T_{K\rm{MB}},
	\label{eq:trans_KM}
\end{equation}
\end{subequations}
\begin{subequations}\label{eq:P_eps_T_WM}
\begin{equation}
		P_{W\rm{M}}
		= + {\cal{R}}^{ab} \frac{\partial B^b}{\partial x^a}
		+ {\bf{E}}_{\rm{M}} \cdot \mbox{\boldmath$\Omega$}
		= - P_W,	
	\label{eq:product_WM}
\end{equation}
\begin{equation}
		\varepsilon_{W\rm{M}}
		= (\nu + \lambda) \frac{\partial U^a}{\partial x^b}
			\frac{\partial B^a}{\partial x^b},
	\label{eq:dissip_WM}
\end{equation}
\begin{equation}
		T_{W\rm{M}} = T_{W\rm{MT}} + T_{W\rm{MB}}.
	\label{eq:trans_WM}
\end{equation}
\end{subequations}
Here in the transport-rate terms, $T_{K\rm{MT}}$ and $T_{W\rm{MT}}$ are the genuine transport rates of the mean MHD energy and cross helicity given by
\begin{equation}
		T_{K\rm{MT}}
		= \nabla \cdot \left( {
		- {\bf{U}} : \mbox{\boldmath${\cal{R}}$}
		+ {\bf{E}}_{\rm{M}} \times {\bf{B}}
		} \right),
	\label{eq:T_KMT_def}
\end{equation}
\begin{equation}
		T_{W\rm{MT}}
		= \nabla \cdot \left( {
		- {\bf{B}} : \mbox{\boldmath${\cal{R}}$}
		+ {\bf{E}}_{\rm{M}} \times {\bf{U}}
		} \right)
	\label{eq:T_WMT_def}
\end{equation}
[$({\bf{U}}:\mbox{\boldmath${\cal{R}}$})^\alpha= U^b {\cal{R}}^{b\alpha}$ and $({\bf{B}}:\mbox{\boldmath${\cal{R}}$})^\alpha= B^b {\cal{R}}^{b\alpha}$]. We have dropped the $\nu$- and $\lambda$-related terms in Equations~(\ref{eq:T_KMT_def}) and (\ref{eq:T_WMT_def}). 

	On the other hand, $T_{K\rm{MB}}$ in Equation~(\ref{eq:trans_KM}) and $T_{W\rm{MB}}$ in Equation~(\ref{eq:trans_WM}) are the transport rates arising from the inhomogeneities along the mean magnetic field $\bf{B}$ and from external force. If we have some inhomogeneity along the mean magnetic field, these terms work as a kind of generation mechanism for the large-scale MHD energy and cross helicity. These terms can be written as
\begin{equation}
		T_{K\rm{MB}}
		= {\bf{B}} \cdot \left[ {
			\nabla \left( {
			{\bf{U}} \cdot {\bf{B}}
			} \right)
		} \right]
		+ {\bf{U}} \cdot \left( {{\bf{F}} - \nabla P_{\rm{M}}} \right),
	\label{eq:T_KMB_def}
\end{equation}
\begin{equation}
		T_{W\rm{MB}}
		= {\bf{B}} \cdot \left[ {
			\nabla \left( {
			\frac{{\bf{U}}^2 + {\bf{B}}^2}{2}
			} \right)
			+ {\bf{F} - \nabla P_{\rm{M}}}
		} \right].
	\label{eq:T_WMB_def}
\end{equation}

	We should note that the production rates of the mean MHD energy and cross helicity, $P_{K\rm{M}}$ [Equation~(\ref{eq:product_KM})] and $P_{W\rm{M}}$ [Equation~(\ref{eq:product_WM})], are exactly the same as the counterparts of the turbulent MHD energy and cross helicity, $P_K$ [Equation~(\ref{eq:P_K_def})] and $P_W$ [Equation~(\ref{eq:P_W_def})], respectively, with the reversed sign. This shows the productions of the turbulent MHD energy $\langle {{\bf{u}}'{}^2 + {\bf{b}}'}^2 \rangle/2$ and cross helicity $\langle {{\bf{u}}' \cdot {\bf{b}}'} \rangle$ correspond to the sinks or drains of the large-scale MHD energy $({\bf{U}}^2 + {\bf{B}}^2)/2$ and cross helicity ${\bf{U}} \cdot {\bf{B}}$. In this sense, the evolution of these turbulent quantities are subject to a kind of cascade process in turbulence.

\section{Turbulent cross-helicity expression based on the TSDIA}\label{sec:appendixB}
	In terms of the Elssasser variables
\begin{equation}
	\mbox{\boldmath$\phi$} = {\bf{u}} + {\bf{b}},\
	\mbox{\boldmath$\psi$} = {\bf{u}} - {\bf{b}},
\end{equation}
the magnetohydrodynamic (MHD) equations of the incompressible fluid can be written as
\begin{equation}
	\frac{\partial \mbox{\boldmath$\phi$}}{\partial t}
	+ (\mbox{\boldmath$\psi$} \cdot \nabla)
	\mbox{\boldmath$\phi$}
	= - \nabla p_{\rm{M}}
	+ \nu \nabla^2 \mbox{\boldmath$\phi$},
	\label{eq:elss_phi_eq}
\end{equation}
\begin{equation}
	\nabla \cdot \mbox{\boldmath$\phi$} = 0,
	\label{eq:elss_phi_sol}
\end{equation}
and the counterparts obtained by the interchange of $\mbox{\boldmath$\phi$} 
\leftrightarrow \mbox{\boldmath$\psi$}$. Here, $p_M (= p + {\bf{b}}^2/2)$ is the MHD pressure. We have assumed the difference between the viscosity and magnetic diffusivities are small compared to the sum of them and neglect the former, putting $\nu = \eta$. As for the extension of the Elssasser formalism to the compressible case, the reader is referred to \cite{mar1987} (and also \cite{yok2007} in the context of turbulence).

	In the two-scale direct-interaction approximation (TSDIA) formalism, we introduce two-scale variables: the fast and slow variables as
\begin{equation}
	\mbox{\boldmath$\xi$} (= {\bf{x}}),\ {\bf{X}} (= \delta {\bf{x}});\ 
	\tau (= t),\ T (= \delta t).
	\label{eq:fast_slow_vars}
\end{equation}
As this result, the spatial and time derivatives are written as
\begin{equation}
	\nabla = \nabla_{\mbox{\boldmath$\xi$}} + \delta \nabla_{\bf{X}};\
	\frac{\partial}{\partial t} 
	= \frac{\partial}{\partial \tau} + \delta \frac{\partial}{\partial T}.
	\label{eq:diff_exp}
\end{equation}
This shows the property of derivative expansion with respect to the slow variables with $\delta$ being the scale parameter. With the aid of Equation~(\ref{eq:fast_slow_vars}), we separate the fast variation of fluctuation from the slow variation of mean as
\begin{equation}
	f = F({\bf{X}};T) + f'(\mbox{\boldmath$\xi$}, {\bf{X}}; \tau, T).
	\label{eq:two-scl_div}
\end{equation}
	The Elssasser variables are divided into means and fluctuations as
\begin{equation}
	\mbox{\boldmath$\phi$} = \mbox{\boldmath$\Phi$} + \mbox{\boldmath$\phi$}',\
	\mbox{\boldmath$\psi$} = \mbox{\boldmath$\Psi$} + \mbox{\boldmath$\psi$}'.
	\label{eq:elss_divis}
\end{equation}
We substitute Equation~(\ref{eq:elss_divis}) into Equations~(\ref{eq:elss_phi_eq}) and (\ref{eq:elss_phi_sol}) in a rotating system with the angular velocity of $\mbox{\boldmath$\Omega$}_{\rm{F}}$. With Equations~(\ref{eq:fast_slow_vars})-(\ref{eq:two-scl_div}), the governing equation for the lowest-order fields in $\delta$ is written in the wavenumber space as
\begin{eqnarray}
	\lefteqn{
	\frac{\partial\phi_{0}'^a({\bf{k}};\tau)}{\partial\tau}
	+ \nu k^2 \phi_{0}'^a({\bf{k}};\tau)
	}\nonumber\\
	& & - i Z^{abc}({\bf{k}}) \iint \delta({\bf{k}}-{\bf{p}}-{\bf{q}}) d{\bf{p}} d{\bf{q}}\
	\psi_{0}'^b({\bf{p}};\tau)
	\phi_{0}'^c({\bf{q}};\tau)
	\nonumber\\
	& & = - i ({\bf{k}}\cdot{\bf{B}}) \phi_0'^a({\bf{k}};\tau)
	- \epsilon^{dbc} \Omega_{\rm{F}}^b D^{ad}({\bf{k}}) 
		\left( {\phi_0'^c({\bf{k}};\tau) \psi_0'^c({\bf{k}};\tau)} \right),
	\label{eq:phi0_eq}
\end{eqnarray}
and the $\mbox{\boldmath$\psi$}$ counterpart. Here, $\bf{B}$ is the mean magnetic field, $D^{ab}({\bf{k}}) (= \delta^{ab} - k^a k^b /k^2)$ the projection operator, and $Z^{abc}({\bf{k}}) = k^a D^{bc}({\bf{k}})$. Hereafter we suppress the dependence on the slow variables, $\bf{X}$ and $T$, except when necessary. In Equation~(\ref{eq:phi0_eq}), the MHD-pressure term has been eliminated by using the Poisson equation for it.

	The equation for the first-order field is given as
\begin{eqnarray}
	\lefteqn{
	\frac{\partial\phi_{\rm{S}1}'^a({\bf{k}};\tau)}{\partial\tau}
	+ \nu k^2 \phi_{\rm{S}1}'^a({\bf{k}};\tau)
	}\nonumber\\
	& & - i Z^{abc}({\bf{k}}) \iint \delta({\bf{k}}-{\bf{p}}-{\bf{q}}) d{\bf{p}} d{\bf{q}}\
	\psi_{0}'^b({\bf{p}};\tau)
	\phi_{\rm{S}1}'^c({\bf{q}};\tau)
	\nonumber\\
	& & = - D^{ac}({\bf{k}}) \psi_0'^b({\bf{k}};\tau) 
		\frac{\partial \Phi^c}{\partial X^b}
	- \frac{D\phi_0'^a}{DT_{\rm{I}}}
	+ B^b \frac{\partial \phi_0'^a({\bf{k}};\tau)}{\partial X_{\rm{I}}^b}
	\nonumber\\
	& & + i \epsilon^{dbc} \Omega_{\rm{F}}^b 
		\frac{k^c}{k^2} D^{ad}({\bf{k}}) 
		\frac{\partial}{\partial X_{\rm{I}}^e}
	\left( {
		\phi_0'^e({\bf{k}};\tau) 
		+ \psi_{0}'^e({\bf{k}};\tau)
	} \right)
	\nonumber\\
	& & + i \epsilon^{dbc} \Omega_{\rm{F}}^b 
		\frac{k^d}{k^2} D^{ae}({\bf{k}}) 
		\frac{\partial}{\partial X_{\rm{I}}^e}
	\left( {
		\phi_0'^c({\bf{k}};\tau) 
		+ \psi_{0}'^c({\bf{k}};\tau)
	} \right)
	\nonumber\\
	& & - i ({\bf{k}}\cdot{\bf{B}}) \phi_{\rm{S}1}'^a({\bf{k}};\tau)
	- \epsilon^{dbc} \Omega_{\rm{F}}^b D^{ad}({\bf{k}}) \left( {
		\phi_{\rm{S}1}'^c({\bf{k}};\tau) 
		+ \psi_{\rm{S}1}'^c({\bf{k}};\tau)
	} \right).
	\label{eq:phiS1_eq}
\end{eqnarray}
Here, $\mbox{\boldmath$\phi$}'_{\rm{S}}$ is the first-order field satisfying the solenoidal condition [See Equation~(\ref{eq:phi_S1_def}) later]. The $\mbox{\boldmath$\psi$}$ counterpart can be obtained by exchanging
\begin{equation}
	\mbox{\boldmath$\phi$} \leftrightarrow \mbox{\boldmath$\psi$},\
	\mbox{\boldmath$\Phi$} \leftrightarrow \mbox{\boldmath$\Psi$},\
	\mbox{\boldmath$\phi$}' \leftrightarrow \mbox{\boldmath$\psi$}',\
	{\bf{B}} \leftrightarrow -{\bf{B}}
\end{equation}
in Equation~(\ref{eq:phiS1_eq}).

	In Equation~(\ref{eq:phi0_eq}) we consider the turbulence fields that is free from factors generating the anisotoropy of fluctuations such as the rotation and the magnetic field, and denote them  the basic fields as $\mbox{\boldmath$\phi$}'_{\rm{B}}$ and $\mbox{\boldmath$\psi$}'_{\rm{B}}$. They obey equations similar to the homogeneous isotropic turbulence as
\begin{eqnarray}
	\lefteqn{
	\frac{\partial\phi_{\rm{B}}'^a({\bf{k}};\tau)}{\partial\tau}
	+ \nu k^2 \phi_{\rm{B}}'^a({\bf{k}};\tau)
	}\nonumber\\
	& & - i Z^{abc}({\bf{k}}) \iint \delta({\bf{k}}-{\bf{p}}-{\bf{q}}) d{\bf{p}} d{\bf{q}}\
	\psi_{\rm{B}}'^b({\bf{p}};\tau)
	\phi_{\rm{B}}'^c({\bf{q}};\tau)
	= 0.
	\label{eq:basic_fld_eq}
\end{eqnarray}
Corresponding to this equation, we introduce the Green's function
\begin{eqnarray}
	\lefteqn{
	\frac{\partial G_{\phi}'^{ab}({\bf{k}};\tau,\tau')}
		{\partial \tau}
	+ \nu k^2 G_{\phi}'^{ab}({\bf{k}};\tau,\tau)
	}\nonumber\\
	& & \hspace{-11pt}
	- i Z^{acd}({\bf{k}}) \iint \delta({\bf{k}}-{\bf{p}}-{\bf{q}}) d{\bf{p}} d{\bf{q}}\
	\psi_{0}'^c({\bf{p}};\tau)
	G_{\phi}'^{db}({\bf{q}};\tau,\tau')
	= \delta^{ab} \delta(\tau - \tau').
	\label{eq:green_fn_eq}
\end{eqnarray}
Using this Green's function, we formally solve the first-order field as
\begin{equation}
	\mbox{\boldmath$\phi$}'_1({\bf{k}};\tau)
	= \mbox{\boldmath$\phi$}'_{\rm{S}1}({\bf{k}};\tau)
	- i \frac{\bf{k}}{k^2}
		\frac{\partial \phi_{\rm{B}}'^{a}({\bf{k}};\tau)}{\partial X_{\rm{I}}^a}
	\label{eq:phi_S1_def}
\end{equation}
with
\begin{eqnarray}
	\lefteqn{
	\phi_{\rm{S}1}'^a({\bf{k}};\tau)
	= - \frac{\partial \Phi^c}{\partial X^b}
	D^{cd}({\bf{k}}) \int_{-\infty}^{\tau} \!\!\!d\tau_1\
		G_{\phi}'^{ad} ({\bf{k}};\tau,\tau_1)
		\psi_{\rm{B}}'^b({\bf{k}};\tau_1)
	}\nonumber\\
	& &  - \int_{-\infty}^{\tau} \!\!\!d\tau_1\
		G_{\phi}'^{ab}({\bf{k}};\tau,\tau_1)
		\frac{D\phi_{\rm{B}}'^b({\bf{k}};\tau_1)}{DT_{\rm{I}}}
	\nonumber\\
	& & + B^b \int_{-\infty}^{\tau} \!\!\!d\tau_1\
		G_{\phi}'^{ac}({\bf{k}};\tau,\tau_1)
		\frac{\partial \phi_{\rm{B}}'^c({\bf{k}};\tau_1)}	{\partial X_{\rm{I}}^b}
	\nonumber\\
	& & + i \epsilon^{dbc} \Omega_{\rm{F}}^b 
		\frac{k^c}{k^2} D^{df}({\bf{k}}) 
		\int_{-\infty}^{\tau} \!\!\!d\tau_1\
		G_{\phi}'^{af} ({\bf{k}};\tau,\tau_1)
		\frac{\partial}{\partial X_{\rm{I}}^e}
		\left[ { 
			\phi_{\rm{B}}'^e({\bf{k}};\tau_1) 
			+ \psi_{\rm{B}}'^e({\bf{k}};\tau_1)
		} \right]
	\nonumber\\
	& & + i \epsilon^{dbc} \Omega_{\rm{F}}^b 
		\frac{k^d}{k^2} D^{ae}({\bf{k}})
		\int_{-\infty}^{\tau} \!\!\!d\tau_1\
		G_{\phi}'^{ab} ({\bf{k}};\tau,\tau_1)
		\frac{\partial}{\partial X_{\rm{I}}^e}
	\left[ {
		\phi_{\rm{B}}'^c({\bf{k}};\tau_1) 
		+ \psi_{\rm{B}}'^c({\bf{k}};\tau_1)
	} \right].
	\label{eq:formal_sol_phi_S1}
\end{eqnarray}

	The turbulent cross helicity is defined by
\begin{equation}
	W = \left\langle {
	{\bf{u}}' \cdot {\bf{b}}'
	} \right\rangle
	= \langle { (\mbox{\boldmath$\phi$}'^2
		- \mbox{\boldmath$\psi$}'^2) /4 } \rangle.
	\label{eq:elss_crss_hel_def}
\end{equation}
Multiplying Equation~(\ref{eq:phi_S1_def}) with Equation~(\ref{eq:formal_sol_phi_S1}) by $\mbox{\boldmath$\phi$}'({\bf{k}};\tau)$ and averaging, we integrate it over ${\bf{k}}$. Then we obtain the expression for the turbulent cross helicity as
\begin{eqnarray}
	\lefteqn{
	W 
	= 2 I_0 \{Q_{ub}\}
	}\nonumber\\
	& & - \left[ {
		I_0 \left\{ {
			G_{\rm{S}}, \frac{D}{Dt} \left( {
			Q_{ub} + Q_{bu}
			} \right)
		} \right\}
	+ I_0 \left\{ {
			G_{\rm{A}}, \frac{D}{Dt} \left( {
			Q_{uu} + Q_{bb}
			} \right)
		} \right\}
	} \right]
	\nonumber\\
	& & + \frac{1}{3} \left( {
			\mbox{\boldmath$\Omega$} 
			+ 2 \mbox{\boldmath$\Omega$}_{\rm{F}}
		} \right) \cdot \left[ {
		I_{-1}\left\{ {
		G_{\rm{S}}, \nabla H_{bu}
		} \right\}
		+ 	I_{-1}\left\{ {
		G_{\rm{A}}, \nabla H_{uu}
		} \right\}
	} \right]
	\nonumber\\
	& & - {\bf{B}} \cdot \left[ {
		I_{0}\left\{ {
		G_{\rm{S}}, \nabla \left( {Q_{uu} + Q_{bb}} \right)
		} \right\}
	- 	I_{0}\left\{ {
		G_{\rm{A}}, \nabla \left( {Q_{ub} + Q_{bu}} \right)
		} \right\}
	} \right],
	\label{eq:W_exprss_upto2}
\end{eqnarray}
where use has been made of the abbreviated form of integrals [Equation~(\ref{eq:abbrevs})]. Here we have introduced the statistically isotropic Green's function
\begin{equation}
	\left\langle {
		G_\vartheta^{ab}({\bf{k}};\tau,\tau')
	} \right\rangle
	= \delta^{ab} G_\vartheta(k;\tau,\tau')
\end{equation}
with $\vartheta = (\phi, \psi)$, and defined the mirrorsymmetric and anti-mirrorsymmetric parts of $G_\phi$ and $G_\psi$, $G_{\rm{S}}$ and $G_{\rm{A}}$, by
\begin{subequations}
\begin{equation}
	G_{\rm{S}}(k;\tau,\tau') 
	= (1/2) \left[ {
		G_\phi(k;\tau,\tau') + G_\psi(k;\tau,\tau') 
	} \right],
\end{equation}
\begin{equation}
	G_{\rm{A}}(k;\tau,\tau') 
	= (1/2) \left[ {
		G_\phi(k;\tau,\tau') - G_\psi(k;\tau,\tau') 
	} \right].
\end{equation}
\end{subequations}

\section{Suggestions from higher-order terms}\label{sec:appendixC}
	We consider situations where the time scales are mirrorsymmetric. In such a case, the $G_{\rm{A}}$-related terms may be dropped. If we retain only the $G_{\rm{S}}$-related terms in Equation~(\ref{eq:W_exprss_upto2}), $W$ is expressed as
\begin{eqnarray}
	\lefteqn{
	W 
	= 2 I_0 \{Q_{ub}\}
	- I_0 \left\{ {
		G_{\rm{S}}, \frac{D}{Dt} \left( {
		Q_{ub} + Q_{bu}
		} \right)
	} \right\}
	}\nonumber\\
	& & + \frac{1}{3} \left( {
		\mbox{\boldmath$\Omega$} 
		+ 2 \mbox{\boldmath$\Omega$}_{\rm{F}}
	} \right) \cdot
	I_{-1}\left\{ {
	G_{\rm{S}}, \nabla H_{bu}
	} \right\}
	- {\bf{B}} \cdot
	I_{0}\left\{ {
	G_{\rm{S}}, \nabla \left( {Q_{uu} + Q_{bb}} \right)
	} \right\}.
	\label{eq:W_only_S}
\end{eqnarray}
Here, $Q_{uu}$, $Q_{bb}$, and $H_{bu}$ are the spectral functions of the turbulent quantities at the lowest order, which are defined in terms of the basic or lowest-order fields ${\bf{u}}'_{\rm{B}}$ and ${\bf{b}}'_{\rm{B}}$ as
\begin{equation}
	\frac{1}{2} \left\langle {
	{\bf{u}}'_{\rm{B}}{}^2 + {\bf{b}}'_{\rm{B}}{}^2
	} \right\rangle
	= 2 \int { \left[ {
		Q_{uu}(k;\tau,\tau) + Q_{bb}(k;\tau,\tau)
	} \right] } d{\bf{k}},
\end{equation}
\begin{equation}
	\left\langle {
	{\bf{b}}'_{\rm{B}} \cdot {\mbox{\boldmath$\omega$}}'_{\rm{B}}
	} \right\rangle
	=\int { H_{bu}(k;\tau,\tau)}  d{\bf{k}}.
\end{equation}

	In a manner similar to Section~\ref{sec:level4-2}, we estimate the higher-order or the third and fourth terms in Equation~(\ref{eq:W_only_S}). We assume the spectral functions of the turbulent MHD energy and torsional correlation are written as
\begin{equation}
	Q_{uu}(k,{\bf{x}};\tau,\tau',t)
	+ Q_{bb}(k,{\bf{x}};\tau,\tau',t)
	= \sigma_K(k,{\bf{x}};t)
	\exp\left[ {-\omega_K(k,{\bf{x}};t)|\tau - \tau'|} \right],
	\label{eq:Quu_Qbb_spect}
\end{equation}
\begin{equation}
	H_{bu}(k,{\bf{x}};\tau,\tau',t)
	= \sigma_H(k,{\bf{x}};t)
	\exp\left[ {-\omega_H(k,{\bf{x}};t)|\tau - \tau'|} \right].
\end{equation}
Here, $\sigma_K$ and $\sigma_H$ are the power spectra of the turbulent MHD energy and torsional correlation, respectively, and $\omega_K$ and $\omega_H$ reflect the time scales of fluctuations. (In this Appendix~\ref{sec:appendixC}, the subscript $H$ denotes the cross torsional correlation $H_{bu}$ not the residual helicity.) Note that in Equation~(\ref{eq:Quu_Qbb_spect}) we assumed the time-scale difference between the velocity and magnetic fluctuations is negligible and denote both the frequencies as $\omega_K$. The spectrum and frequency (reciprocal of time scale) of the turbulent MHD energy can be expressed in the form
\begin{equation}
	\sigma_K(k,{\bf{x}};t) = \sigma_{K0} \varepsilon^{2/3}({\bf{x}};t) k^{-11/3},
	\label{eq:mhd_en_spect}
\end{equation}

\begin{equation}
	\omega_{K}(k,{\bf{x}};t)
	= \omega_{K0} \varepsilon^{1/3} k^{2/3}
	= \tau_W^{-1}
	\label{eq:mhd_en_freq}
\end{equation}
($\sigma_{K0}$, $\omega_{K0}$ are numerical factors). This corresponds to the Kolmogorov scaling of the turbulent MHD energy.

	We also assume the spectrum of the torsional correlation obeys a power law, and put
\begin{equation}
	\sigma_H(k,{\bf{x}};t) 
	= \sigma_{H0} \varepsilon^{-1/3}({\bf{x}};t) \varepsilon_H({\bf{x}};t) k^{-11/3},
	\label{eq:crss_tors_spect}
\end{equation}
\begin{equation}
	\omega_{H}(k,{\bf{x}};t)
	= \omega_{H0} \varepsilon_H^{1/3} k^{1/3}
	= \tau_H^{-1}.
	\label{eq:crss_tors_freq}
\end{equation}
This may be a strong assumption, but as far as the spectral indices are concerned, they do not affect the final form of the dissipation-rate equation but only change the model constants. Alternatively, we may consider that a finite torsional correlation arises from the combination of the helicity and cross-helicity correlations such as
\begin{equation}
	\sigma_H(k,{\bf{x}};t) 
	= \sigma_{H0} \varepsilon^{-4/3}({\bf{x}};t) \varepsilon_{Hu}({\bf{x}};t)
	\varepsilon_W({\bf{x}};t) k^{-11/3}
\end{equation}
($\varepsilon_{Hu}$: the helicity dissipation rate). In any case, we need further information on the torsional correlation or helicity spectrum, which so far we can not fully utilized. So, in this work, we assume the simple forms of Equations~(\ref{eq:crss_tors_spect}) and (\ref{eq:crss_tors_freq}), and see the consequence.

	We substitute Equations~(\ref{eq:mhd_en_spect}) and (\ref{eq:crss_tors_spect}) with Equations~(\ref{eq:mhd_en_freq}) and (\ref{eq:crss_tors_freq}) into the higher-order contributions in Equation~(\ref{eq:W_only_S}). Then we have
\begin{eqnarray}
	\lefteqn{
	-{\bf{B}}\cdot I_0 \left\{ {
		G_{\rm{S}}, \nabla \left( {Q_{uu} + Q_{bb}} \right)
	} \right\}
	}\nonumber\\
	& & = {\bf{B}} \cdot \int d{\bf{k}} 
	\int_{-\infty}^{\tau_1} \!\!\!d\tau_1 G_{\rm{S}}(k,{\bf{x}};\tau,\tau_1,t)
	\nabla \left[ {
		Q_{uu}(k,{\bf{x}};\tau,\tau_1,t)
		+ Q_{bb}(k,{\bf{x}};\tau,\tau_1,t)
	}\right]
	\nonumber\\
	& & = - B^a \int d{\bf{k}} \left[ {
		\frac{1}{\omega_{\rm{S}} + \omega_{K}}
		\frac{\partial \sigma_K(k,{\bf{x}};t)}
				{\partial x^a}
		- \frac{\sigma_{K}(k,{\bf{x}};t)}
				{(\omega_{\rm{S}} + \omega_K)^2}
		\frac{\partial \omega_K(k,{\bf{x}};t)}{\partial x^a}
	} \right],
	\label{eq:B-term_def}
\end{eqnarray}
\begin{eqnarray}
	\lefteqn{
	\frac{1}{3} 
		(\mbox{\boldmath$\Omega$} 
		+ 2 \mbox{\boldmath$\Omega$}_{\rm{F}}) 
		\cdot I_{-1} \left\{ {
		G_{\rm{S}}, \nabla H_{bu}
		} \right\}
		}\nonumber\\
	& & = \frac{1}{3} (\Omega^a + 2 \Omega_{\rm{F}}^a) 
			\int k^{-2} d{\bf{k}} \int_{-\infty}^{\tau_1} \!\!\!d\tau_1 
		G_{\rm{S}}(k,{\bf{x}};\tau,\tau_1,t)
		\nabla H_{bu}(k,{\bf{x}};\tau,\tau_1,t)
		\nonumber\\
	& & =  \frac{1}{3} (\Omega^a + 2 \Omega_{\rm{F}}^a)
		\int k^{-2} d{\bf{k}} 
	\nonumber\\
	& & \hspace{20pt} \times \left[ {
	\frac{1}{\omega_{\rm{S}} + \omega_{H}}
	\frac{\partial \sigma_H(k,{\bf{x}};t)}
			{\partial x^a}
	- \frac{\sigma_{H}(k,{\bf{x}};t)}
			{(\omega_{\rm{S}} + \omega_H)^2}
	\frac{\partial \omega_H(k,{\bf{x}};t)}{\partial x^a}
	} \right].
	\label{eq:Omega-term_def}
\end{eqnarray}

	From Equations~(\ref{eq:mhd_en_freq}) and (\ref{eq:crss_tors_freq}), we construct synthesized time scales as
\begin{eqnarray}
	\frac{1}{\tau_{\rm{SK}}}
	&=& \frac{1}{\tau_{\rm{S}}} + \frac{1}{\tau_K}
	= \omega_{\rm{S}}(k,{\bf{x}};t)
	+ \omega_{K}(k,{\bf{x}};t)
	\nonumber\\
	&=&  \left( {
		\omega_{\rm{S}0} + {\omega_{K0}}
	} \right) \varepsilon^{1/3} k^{2/3}
	\equiv \omega_{\rm{sk}} \varepsilon^{1/3} k^{2/3},
	\label{eq:tau_SK_def}
\end{eqnarray}
\begin{eqnarray}
	\frac{1}{\tau_{\rm{SH}}}
	&=& \frac{1}{\tau_{\rm{S}}} + \frac{1}{\tau_H}
	= \omega_{\rm{S}}(k,{\bf{x}};t)
	+ \omega_{H}(k,{\bf{x}};t)
	\nonumber\\
	&=& \omega_{\rm{S}0} \left[ {
	1 
	+ \frac{\omega_{H0}}{\omega_{\rm{S}0}}
	\left( {
		\frac{\varepsilon_H / k}{\varepsilon}
	} \right)^{1/3}
} \right] \varepsilon^{1/3} k^{2/3}
\simeq \omega_{\rm{sh}} 
	\varepsilon^{1/3} k^{2/3}.
	\label{eq:tau_SH_def}
\end{eqnarray}
Namely, for the simplicity of argument, we assume that the time scales of turbulence are determined primarily by the energy transfer rate $\varepsilon$. If we substitute these expressions (\ref{eq:tau_SK_def}) and (\ref{eq:tau_SH_def}) for $\omega_{\rm{S}} + \omega_K$ and $\omega_{\rm{S}} + \omega_H$ into Equations~(\ref{eq:B-term_def}) and (\ref{eq:Omega-term_def}), respectively, we have
\begin{eqnarray}
	\lefteqn{
	-{\bf{B}}\cdot I_0 \left\{ {
		G_{\rm{S}}, \nabla \left( {Q_{uu} + Q_{bb}} \right)
	} \right\}
	}\nonumber\\
	& & = \frac{1}{(2\pi)^{1/3}} 
		\frac{\sigma_{K0}}{\omega_{\rm{sk}}}
		\varepsilon^{1/3} \ell_{\rm{C}}^{4/3} B^a
	\nonumber\\
	& & \hspace{10pt} \times \left[ {
		\left( {
			- \frac{2}{3}
			+ \frac{1}{2} 
			\frac{\omega_{K0}}{\omega_{\rm{sk}}}
		} \right) \frac{1}{\varepsilon} 
		\frac{\partial \varepsilon}{\partial x^a}
	- 	\left( {
			\frac{11}{3}
			+ \frac{\omega_{K0}}{\omega_{\rm{sk}}}
		} \right) \frac{1}{\ell_{\rm{C}}} 
		\frac{\partial \ell_{\rm{C}}}{\partial x^a}
	} \right],
	\label{eq:B-term_ell}
\end{eqnarray}
\begin{eqnarray}
\lefteqn{
\frac{1}{3} 
	(\mbox{\boldmath$\Omega$} 
	+ 2 \mbox{\boldmath$\Omega$}_{\rm{F}}) 
	\cdot I_{-1} \left\{ {
	G_{\rm{S}}, \nabla H_{bu}
} \right\}
}\nonumber\\
& & =  \frac{1}{15 (2\pi)^{7/3}} 
		\frac{\sigma_{H0}}{\omega_{\rm{sh}}}
	\varepsilon^{-2/3} \varepsilon_H 
	\ell_{\rm{C}}^{10/3}
	(\Omega^a + 2\Omega_{\rm{F}}^a) 
	\left\{ {
	- \frac{1}{\varepsilon} 
		\frac{\partial \varepsilon}{\partial x^a}
	}\right.
	\nonumber\\
	& & \hspace{20pt} \left. {
	+ [3 - A_H(\omega_{\rm{S}0},\omega_{H0})] 
		\frac{1}{\varepsilon_H} 
		\frac{\partial \varepsilon_H}{\partial x^a}
	+ [11 + A_H(\omega_{\rm{S}0},\omega_{H0})]
		\frac{1}{\ell_{\rm{C}}} 
		\frac{\partial \ell_{\rm{C}}}{\partial x^a}
	} \right\}
	\label{eq:Omega-term_ell}
\end{eqnarray}
with
\begin{equation}
	A_H(\omega_{\rm{S}0},\omega_{H0})
	\equiv \frac{30}{33 (2\pi)^{1/3}}
	\frac{\omega_{H0}}{\omega_{\rm{sh}}}
	\left( {
		\frac{\varepsilon_H \ell_{\rm{C}}}{\varepsilon}
	} \right)^{1/3}.
	\label{eq:A_H_def}
\end{equation}
Using an algebraic relation between $\ell_{\rm{C}}$ and $K$:
\begin{subequations}
\begin{equation}
	K = 3 (2\pi)^{1/3} \sigma_{K0} \varepsilon^{2/3} \ell_{\rm{C}}^{2/3}
	\label{algeb_rel_K-ell}
\end{equation}
or
\begin{equation}
	\ell_{\rm{C}} = 3^{-3/2} (2\pi)^{-1/2} \sigma_{K0}^{-3/2} \varepsilon^{-1} K^{3/2},
	\label{algeb_rel_ell-K}
\end{equation}
\end{subequations}
we express Equations~(\ref{eq:B-term_ell}) and (\ref{eq:Omega-term_ell}) partly in terms of $K$ instead of $\ell_{\rm{C}}$. Then we have
\begin{eqnarray}
	\lefteqn{
	-{\bf{B}}\cdot I_0 \left\{ {
		G_{\rm{S}}, \nabla \left( {Q_{uu} + Q_{bb}} \right)
	} \right\}
	}\nonumber\\
	& & = \frac{1}{(2\pi)^{1/3}} 
		\frac{\sigma_{K0}}{\omega_{\rm{sk}}}
		\varepsilon^{1/3} \ell_{\rm{C}}^{4/3} B^a
	\nonumber\\
	& & \hspace{20pt} \times
	\left[ {
		\frac{1}{3}\left( {
			1
			+ \frac{1}{2} 
				\frac{\omega_{K0}}{\omega_{\rm{sk}}}
		} \right) \frac{1}{\varepsilon} 
		\frac{\partial \varepsilon}{\partial x^a}
	- 	\frac{1}{6} \left( {
			\frac{11}{3}
			+ \frac{\omega_{K0}}{\omega_{\rm{sk}}}
		} \right) \frac{1}{K} 
		\frac{\partial K}{\partial x^a}
	} \right],
	\label{eq:B-term_K}
\end{eqnarray}
\begin{eqnarray}
	\lefteqn{
	\frac{1}{3} \left( {
	\mbox{\boldmath$\Omega$}
	+ 2 \mbox{\boldmath$\Omega$}_{\rm{F}}
	} \right) \cdot I_{-1} \left\{ {
	G_{\rm{S}}, \nabla H_bu
	} \right\}
	}\nonumber\\
	& & =  \frac{1}{15 (2\pi)^{7/3}} 
	\frac{\sigma_{H0}}{\omega_{\rm{sh}}}  
	\varepsilon^{1/3} \ell_{\rm{C}}^{4/3}
	\left( {
		\frac{\varepsilon_H \ell_{\rm{C}}}{\varepsilon}
	} \right)
	[\ell_{\rm{C}} (\Omega^a + 2\Omega_{\rm{F}}^a)] \nonumber\\
	& & \hspace{10pt} \times \left\{ {
	- [12 + A_H(\omega_{\rm{S0}},\omega_{H0})] 
		\frac{1}{\varepsilon} 
		\frac{\partial \varepsilon}{\partial x^a}
	+ [3 - A_H(\omega_{\rm{S0}},\omega_{H0})] 
		\frac{1}{\varepsilon_H} 
		\frac{\partial \varepsilon_H}{\partial x^a}
	} \right.
	\nonumber\\
	& &	\hspace{30pt} \left. {
	+ \left[ {
		\frac{33}{2} 
		+ \frac{3}{2} A_H(\omega_{\rm{S0}},\omega_{H0})
		} \right]
		\frac{1}{K} \frac{\partial K}{\partial x^a}
	} \right\}.
	\label{eq:Omega-term_K}
\end{eqnarray}

	Equations (\ref{eq:B-term_K}) and (\ref{eq:Omega-term_K}) show that, in addition to the terms in Equation~(\ref{eq:eps_W_eq_final}), the cross-helicity dissipation equation may contain the higher-order terms related to the mean magnetic field ${\bf{B}}$ and the mean vorticity $\mbox{\boldmath$\Omega$}$ (and the angular velocity $\mbox{\boldmath$\Omega$}_{\rm{F}}$) such as
\begin{equation}
	- \frac{\varepsilon}{K} 
	\left( {{\bf{B}} \cdot \nabla} \right) K,\;\;
	\left( {{\bf{B}} \cdot \nabla} \right) \varepsilon,\;\;
	K^{1/2} \left( {\mbox{\boldmath$\Omega$} \cdot \nabla} \right) K,\;\;
	- \frac{K^{3/2}}{\varepsilon} \left( {\mbox{\boldmath$\Omega$} \cdot \nabla} \right) 	\varepsilon.
\end{equation}
Model constants for these terms can be estimated by time scales of turbulence, $\omega_{\rm{S}}$, $\omega_K$, and $\omega_H$. At the same time, inhomogeneity of the cross-torsional correlation $\langle {{\bf{b}}' \cdot \mbox{\boldmath$\omega$}'} \rangle$, which is related to the coupling of the helicity and cross helicity, may play a certain role in the cross-helicity dissipation.

\end{document}